\documentclass[a4paper,12pt]{article}

\usepackage{amsmath}
\usepackage[psamsfonts]{amssymb}
\usepackage{rsfs}
\usepackage{bm}

\usepackage{cite}
\usepackage[dvips]{graphicx}
\usepackage{color}

\makeatletter
\@addtoreset{equation}{section}
\makeatother

\begin{document} 

\begin{titlepage}

\baselineskip 10pt
\hrule 
\vskip 5pt
\leftline{}
\leftline{Chiba Univ. Preprint
          \hfill   \small \hbox{\bf CHIBA-EP-139}}
\leftline{\hfill   \small \hbox{hep-th/0303251}}
\leftline{\hfill   \small \hbox{March 2003}}
\vskip 5pt
\baselineskip 14pt
\hrule 
\vskip 1.0cm
\centerline{\Large\bf 
Implications of Analyticity
} 
\vskip 0.3cm
\centerline{\Large\bf  
to Mass Gap, Color Confinement
}
\vskip 0.3cm
\centerline{\Large\bf  
and Infrared Fixed Point in Yang--Mills theory
}
\vskip 0.3cm
\centerline{\large\bf  
}

\vskip 0.5cm

\centerline{{\bf 
Kei-Ichi Kondo$^{\dagger,{1}}$ 
}}  
\vskip 0.5cm
\centerline{\it
${}^{\dagger}$Department of Physics, Faculty of Science, 
Chiba University, Chiba 263-8522, Japan
}
\vskip 1cm

\begin{abstract}
Analyticity of gluon and Faddeev--Popov ghost propagators and their form factors on the complex momentum-squared plane is exploited to  continue analytically the ultraviolet asymptotic form calculable by perturbation theory into the infrared non-perturbative solution.
We require  the non-perturbative multiplicative renormalizability  to write down the renormalization group equation. 
These requirements enable one to settle the value of the exponent characterizing the infrared asymptotic solution  with power behavior which was originally predicted by Gribov and has recently been found as approximate solutions of the coupled truncated Schwinger--Dyson equations.
For this purpose, we have obtained all the possible superconvergence relations for the propagators and form factors in both the generalized Lorentz gauge and the modified Maximal Abelian gauge.  We show that the transverse gluon propagators are suppressed in the infrared region to be of the massive type irrespective of the gauge parameter, in agreement with the recent result of numerical simulations on a lattice.  
However, this method alone is not sufficient to specify some of the ghost propagators which play the crucial role in color confinement. 
Combining the above result with the renormalization group equation again, we find an infrared enhanced asymptotic solution for the ghost propagator.  The coupled solutions  fulfill the color confinement criterion due to Kugo and Ojima and also Nishijima, at least, in the Lorentz--Landau gauge. 
We also point out that the solution in compatible with color confinement leads to the existence of the infrared fixed point in pure Yang--Mills theory without dynamical quarks. 
Finally, the Maximal Abelian gauge is also examined in connection with quark confinement.

\end{abstract}


\hrule  
\vskip 0.1cm
${}^1$ 
  E-mail:  {\tt kondok@faculty.chiba-u.jp}

\par 
\par\noindent


\vskip 0.5cm

\newpage
\pagenumbering{roman}
\tableofcontents

\vskip 0.5cm  



\end{titlepage}


\pagenumbering{arabic}

\baselineskip 14pt
\section{Introduction}

Quantum chromodynamics (QCD), a non-Abelian gauge theory \cite{YM54} with a gauge group $SU(3)$, is now regarded as the fundamental theory for describing strong interactions among quarks, gluons and hadrons as their bound states. 
Since  asymptotic freedom \cite{AF} exists in QCD with $N_f$ quark flavors less than 33/2, 
one can predict the ultraviolet (UV) behavior of QCD by perturbation theory in a reliable way.  
In order to discuss the infrared (IR) behavior, on the other hand, we must solve the strong coupling problem beyond the perturbation theory.  In the IR region, indeed, the perturbation theory itself cannot be a reliable method of calculations.  
Hence, we need inevitably non-perturbative methods. 
However, this does not necessarily mean that 
the perturbation theory cannot provide any information on the IR behavior of QCD, as will be shown in this article.

One of the most direct methods to study the non-perturbative features of  quantum field theory would be to write down and solve the Schwinger--Dyson (SD) equation.  The SD equation is a set of self-consistent equations among Green functions and constitutes a series of exact equations.   
However, the SD equations consist of the infinite number of simultaneous non-linear integral equations.
In actually solving the SD equation, therefore, one needs to perform a certain truncation which can not be justified within the framework. 
Nevertheless, the approximate solutions have become an efficient tool for studying non-perturbative problems which cannot be reached by perturbative method, e.g. chiral symmetry breaking. See e.g., reviews \cite{RW94,RS00,AS01}.

\begin{figure}[htbp]
\begin{center}
\includegraphics[height=14cm]{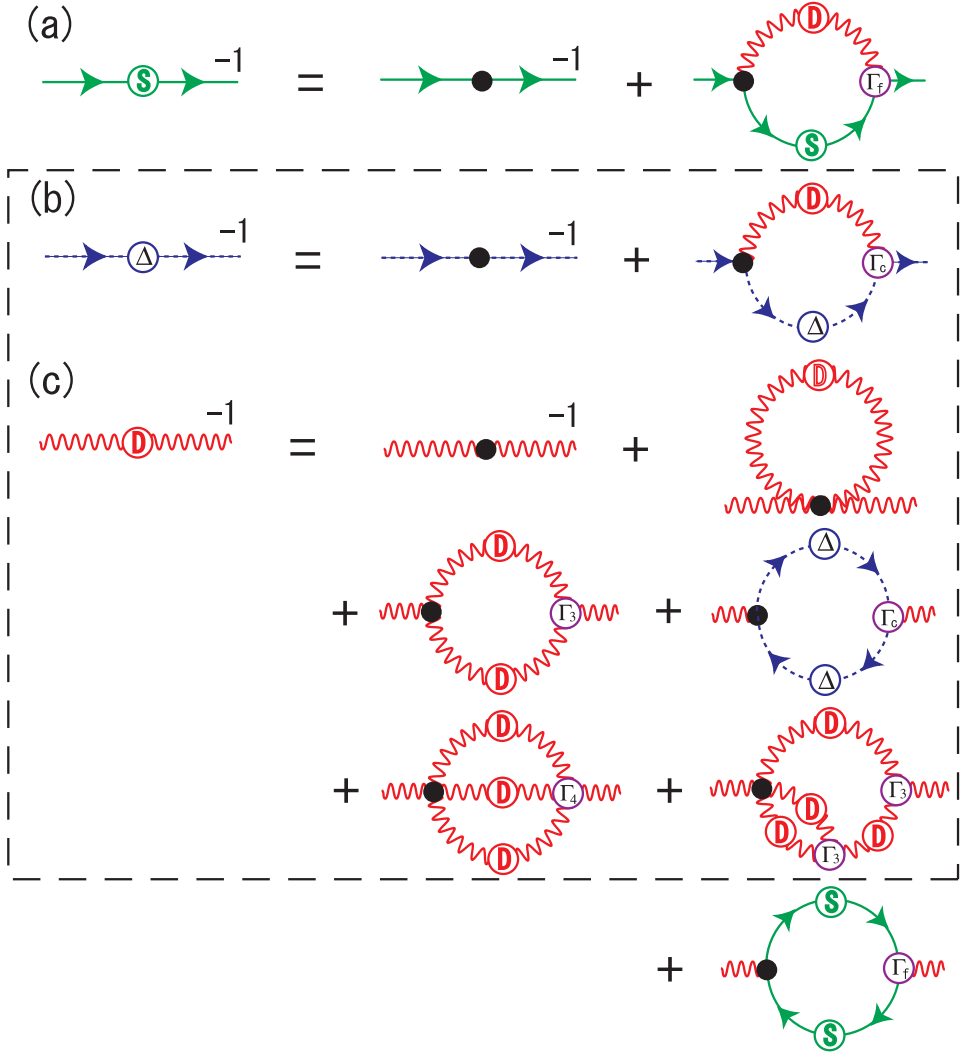}
\caption{The coupled Schwinger--Dyson equations for three propagators in QCD with  the conventional Lorentz gauge fixing.
 (a) quark equation, (b) ghost equation, (c) gluon equation.
Here $S$ denotes the full quark propagator, $\Delta$ the full ghost propagator and $D$ the full gluon propagator, while $\Gamma$ denote the four types of vertices. 
The pure Yang-Mills part is enclosed by the broken line. 
}
\label{fig:QCD-YM_SDeq}
\end{center}
\end{figure}

So far, the extensive studies of the SD equations in  QCD have  almost been restricted to the SD equation for quark propagator (referred to as quark SD equation hereafter), see Figure~\ref{fig:QCD-YM_SDeq}(a).  
In order to study gluodynamics,%
\footnote{In this article, we deal only with the manifestly Lorentz covariant formulation of Yang--Mills theory. Therefore, we do not discuss the non-covariant gauge, such as the Coulomb gauge and axial gauge \cite{BZ99}.  We do not discuss the Fock--Schwinger gauge in this article either. 
}
we must solve the coupled SD equations for the gluon and Faddeev--Popov  (FP) \cite{FP67} ghost propagators, see Figure~\ref{fig:QCD-YM_SDeq}(b),(c).  Of course, solving them is more difficult than solving the quark equation alone.  This is mainly because the SD equations for gluon and ghost are {\it truly non-linear} coupled equations.  
This is in sharp contrast with the quark SD equation studied so far.

An origin of this non-linearity comes from the treatment in which both propagators are incorporated on equal footing, see Figure~\ref{fig:QCD-YM_SDeq}(b),(c). 
In the conventional treatment of the quark SD equation, the gluon propagator was replaced by hand with the free propagator (ladder approximation) or  dressed propagator in the one-loop resumed level (improved ladder approximation) and consequently the problem was reduced to solving  the quark propagator alone.%
\footnote{For the four-dimensional QED, the coupled SD equations for electron propagator and photon propagator in the Landau gauge were solved a decade ago under the bare vertex approximation in an article \cite{KMN92} and subsequently by Bloch and Pennington \cite{BP95} in a numerical way. 
}
Another origin of non-linearity comes from the fact that gluon and ghost are bosons. 
The quark SD equation can be decomposed into two equations for the mass function and the wave-function renormalization function.
The equation for the quark mass function can be linearized as a good  approximation or in the rigorous sense (bifurcation technique) due to a fact that quark is a fermion.   
The {\it general solution} for the mass function can be obtained 
by adopting either method of linearizations.  
Once the general solution is obtained, it can be used to study the whole range of momenta.  This fact facilitates considerably the analysis of the quark mass function.


\par
In recent several years,  extensive analytical studies \cite{SHA97,AB98,AB98b,AAS99,WA01,Ahlig01,Bloch01,Bloch02,FAR02,LS02,Kondo02a,Kondo02b,FS03,Bloch03} of the coupled SD equations for gluon and ghost propagators were performed to obtain the  infrared (IR) and ultraviolet (UV) asymptotic solutions in the  $SU(N_c)$ Yang--Mills theory restricted to the Landau gauge.  
The leading term of the UV asymptotic solution recovers the one-loop resumed perturbation theory  due to asymptotic freedom.  In fact, the solutions for the Euclidean gluon propagator 
$
 D_{\mu\nu}^{AB}(Q^2) = \delta^{AB} (\delta_{\mu\nu}-Q_\mu Q_\nu/Q^2)D_T(Q^2)
$
and  ghost propagator 
$\Delta_{FP}^{AB}(Q^2)=\delta^{AB} \Delta_{FP}(Q^2)$
($A,B=1, \cdots, N_c^2-1$ for $SU(N_c)$)
have the expected UV forms
\begin{align}
  D_T(Q^2) \cong {C \over Q^2} (\log Q^2)^\gamma , \quad \Delta_{FP}(Q^2) \cong {D \over Q^2} (\log Q^2)^\delta ,
\label{UVsol}
\end{align}
where $C$ and $D$ are $Q^2$-independent constants and  the exponents $\gamma$ and $\delta$ are independent of the number of colors $N_c$ in $SU(N_c)$ Yang-Mills theory at one-loop level, i.e., 
$\gamma=-{13 \over 22}$ and $\delta=-{9 \over 44}$.

A remarkable characteristic of gluodynamics claimed by von Smekal, Hauck and Alkofer \cite{SHA97} and subsequently confirmed by Atkinson and Bloch \cite{AB98} in a simpler framework is 
that the Euclidean gluon propagator 
behaves like a power of the momenta in the IR limit of vanishing momenta $Q^2 \rightarrow 0$:
\begin{equation}
  D_T(Q^2) \cong A (Q^2)^{2\kappa-1} ,
\label{Dsol}
\end{equation}
and the gluon propagator is suppressed in the IR regime for $\kappa>0$, 
while  the Euclidean ghost propagator is enhanced  to be more singular than the free one in the IR limit $Q^2 \rightarrow 0$:
\begin{equation}
 \Delta_{FP}(Q^2) \cong B (Q^2)^{-\kappa-1}  ,
\label{Gsol}
\end{equation}  
where $A$ and $B$ are $Q^2$-independent constants.
This result is surprisingly due to dominant contribution from the ghost loop in IR region.
However, the value of the exponent $\kappa$ varies in the range 
$0<\kappa \le 1$  depending on the way of truncation  
\cite{AB98,AB98b,AAS99,WA01,Ahlig01,Bloch01,Bloch02,FAR02,LS02,Kondo02a,Kondo02b,FS03,Bloch03}.
At present, the precise value of $\kappa$ is not known. 
Therefore, it is still under the debate whether the gluon propagator converges to a finite non-zero value ($\kappa=1/2$) or vanishes ($\kappa>1/2$) in the IR limit $Q^2 \rightarrow 0$.%
\footnote{On the lattice, there is a general argument \cite{Zwanziger91} which supports the vanishing of the gluon propagator at $Q=0$.  However, this is not necessarily the case in the continuum limit due to the renormalization effect, see pp.146--147 of \cite{Zwanziger91}. 
}
It should be remarked that such behaviors of gluon and ghost propagators were first predicted by Gribov \cite{Gribov78}. His result corresponds to $\kappa=1$:
\begin{equation}
  D_T(Q^2) \sim {Q^2 \over Q^4+M_G^4}  , \quad 
 \Delta_{FP}(Q^2) \sim {M_G^2 \over (Q^2)^{2}}  ,
\end{equation}
where the mass scale $M_G$ is called the Gribov scale.

\par
A serious problem is that at present one does not have any {\it analytical method} for connecting in the intermediate region two asymptotic solutions to obtain the complete solution over the whole range of momenta.  Only the numerical studies have succeeded in obtaining the complete solution in this sense.
Therefore, there is no guarantee that a specific IR asymptotic solution obtained in the analytical study can be a piece of the complete solution which must reduce to the UV asymptotic solution in the large momenta. 
There is also a problem of uniqueness of the solution \cite{Kondo02a}, since the IR asymptotic solution is obtained based on a specific Ansatz for the solution.  Therefore, one cannot judge if a set of IR power solutions (\ref{Dsol}) and (\ref{Gsol}) is unique for the coupled SD equations. 
The other problem is that all the SD equations aforementioned have been solved in Euclidean space. Indeed, the obtained Euclidean solutions are very helpful for  comparison with the results of numerical simulations on a lattice. However, one needs to perform analytic continuation in order to go back to the Minkowski spacetime.  
The analytic continuation to the Minkowski region is not necessarily  ensured for the Euclidean solutions with  the  asymptotic behaviors (\ref{Dsol}) and (\ref{Gsol}).

The main purpose of this article is to study the relationship between   quark, gluon and ghost propagators in Euclidean region and the spectrum of QCD in the Minkowski region in connection with color confinement.  
In this article, we show in Yang--Mills theory with the gauge group $SU(N_c)$ that the {\it analyticity} of the full propagator on the complex momentum plane plays the role of connecting the perturbative UV asymptotic solution into the non-perturbative IR asymptotic solution and vice versa. 
This study resolves the question as to what the IR asymptotic solution with power behavior (characterized by an exponent $\kappa$) means for the spectrum with respect to the fundamental fields.    
We suggests which type of IR asymptotic solutions should be searched for in the SD equation, while the UV asymptotic solution can be reliably predicted by perturbation theory due to asymptotic freedom.   In other words, the precise value of $\kappa$ will be predicted, if the solution behaves like a power of the momenta. 

In this article, we examine two types of manifestly Lorentz covariant gauges: first the Lorentz gauge in the most general form \cite{CF76,BT82,Baulieu85,DJ82}, so called the Curci--Ferrari gauge, and then the modified version 
\cite{KondoII,KondoIV,KondoV,KondoVI,KS01,SIK01,Kondo01gakkai,EW02,Schaden99,KS00,DV03,Sawayanagi03,EW03} of the Maximal Abelian (MA) gauge \cite{tHooft81,KLSW87}. 
In the conventional Lorentz gauge, all the approximate solutions of the coupled SD equations for the gluon and ghost propagators have been obtained only in the Landau gauge. 
The SD equation approach in the MA gauge in the same level as the Lorentz gauge is still missing in the MA gauge, but see \cite{KMS03} for a preliminary result. 

This article is organized as follows. 
In section 2, we introduce the generalized Lorentz gauge and the modified MA gauge for later convenience. 
For the above purpose, we analyze the propagators in the complex $k^2$ (momentum squared) plane in section 3.  
We require {\it non-perturbative multiplicative renormalizability} for writing down the renormalization group (RG) equation for the propagators in the complex $k^2$ plane.
The  {\it analyticity} of the propagators except for the positive real axis and the asymptotic behavior for large $|k^2|$ enables us to specify the  IR asymptotic form ($|k^2| \rightarrow 0$) such that it can be analytically continued to the UV asymptotic form ($|k^2| \rightarrow \infty$) calculable by perturbation theory.

In this article, 
we investigate which type of IR asymptotic solution of the SD equations is consistent with the general principles of  quantized  gauge field theories:
\begin{enumerate}
\item Non-perturbative multiplicative renormalizability
\item Analyticity
\item Spectral condition (and completeness)
\item Poincar\'e group structure
\end{enumerate}
 without relying on the approximate solutions of the SD equation.

The former half of this article is deeply indebted to an idea of old  articles by Oehme and Zimmerman \cite{OZ80a,OZ80b} and subsequent articles by Oehme and his collaborator \cite{Oehme87,Oehme89,Oehme90a,Oehme90b,OX94,Oehme95a,Oehme95b,OX96,Xu96,Oehme97}. 
 A concept of the superconvergence is  crucially important also in  this article.  The superconvergence relation is nothing but a sum rule for the spectral function of the {\it fundamental field} propagator which is defined in the Minkowski region.%
\footnote{
The various sum rules were devised in QCD to investigate the hadronic properties, see \cite{CK01,Rafael99,Shifman99} for reviews. The sum rule in this article is restricted to that for the fundamental field, rather than the composite hadronic field operators. 
}  
 We reconsider anew the relationship between the superconvergence relation and the Euclidean propagator, which is to be compared with the solution of the coupled SD equations,  paying special attention to the IR asymptotic solution.  
We explicitly obtain all the possible superconvergence relations in the generalized Lorentz gauge in section 4 and the modified MA gauge in section 7. 
In the generalized Lorentz gauge, we show that the superconvergence relation holds  for gluon.  Consequently, 
{\it the Euclidean gluon propagator has the power-series expansion in $Q^2$ and goes to a $Q^2$-independent constant, i.e.,   
$D_T(Q^2) \sim \text{const}.$  
 in the IR limit $Q^2 \rightarrow 0$.
 } 
Therefore,  this result suggests that {\it the IR exponent $\kappa$ characterizing the Euclidean solution must take the value 
\begin{equation}
 \kappa=\frac{1}{2}    
\end{equation}
irrespective of the value of gauge fixing parameter}.

A question is whether or not the IR asymptotic solution for the ghost propagator exhibits power behavior, just as the recent approximate solution of the truncated SD equation suggests.  
So far, one has payed little attention to the superconvergence for  ghost. 
We claim that {\it the superconvergence does not hold for ghost  in the pure Yang--Mills theory (without dynamical fermions)} against the previous result \cite{Xu96}. 
In fact, we elucidate that non-existence of the superconvergence for ghost  is  consistent with realizing  color confinement in QCD.
At the same time, we point out  the subtleties of using the Landau gauge in the Lorentz type gauge fixing, although all the approximate solutions of the coupled SD equations for the gluon and ghost propagators have been obtained only in the Landau gauge. 
In this article we shall provide a unified view among  Euclidean solutions of the SD equations, color confinement and the spectrum, especially the mass gap of Yang--Mills theory and QCD.

In section 5,  the IR asymptotic form in Euclidean region ($k^2 <0$: negative real axis) is related to the spectral property in the Minkowski region ($k^2>0$: positive real axis) by way of the spectral representation  which follows from the spectral condition, Poincar\'e covariance and completeness of the intermediate states.  
We show that {\it $D_T(0)<\infty$ is consistent with the existence of a massive pole in the gluon propagator in the Minkowski region.}%
\footnote{
It may happen that this constant $D_T(0)$ vanishes. 
This case implies $\kappa=1$, the Gribov limit.
This possibility can not be excluded from the general considerations only. However, such a case is realized only when the subtle cancellation occurs between two terms. 
The argument in more detail will be given later. 
}   
Moreover,  we show that, for gluon with  massive spectrum, the gluon propagator and the gluon form factor have the well-defined power-series expansions in $Q^2$ for small Euclidean momenta $Q^2$, suggesting the uniqueness of the IR asymptotic solution with power behavior. 
In the Landau gauge, this result does not contradicts with the existence of the gluon  condensate $\langle A_\mu^2 \rangle$ of mass dimension two which has been extensively studied by several groups \cite{GSZ01,Boucaudetal00,VKAV01,DVBG03}, 
see \cite{Kondo03b} for more details on the physical meaning of dimension two condensate.
The gluon condensate $\langle A_\mu^2 \rangle$  is a special limit of the ghost--gluon condensate of mass dimension 2 which is on-shell BRST invariant as proposed in \cite{Kondo01,KMSI02} and calculated recently in \cite{DVLSS03}.

In the latter half of this article, we shall focus on the following issues. 
For the FP {\it  ghost propagator}, however, the analyticity approach alone is not sufficient to provide precise form of the IR asymptotic solution.
This is reasonable for our result to be consistent with the result of SD equation, since the superconvergence for ghost in the Landau gauge contradicts with the solution of the SD equations mentioned above, i.e., enhancement in the IR region. In section 6, therefore, we make use of the  RG equations for the gluon and ghost propagators in the IR region.
First, we point out that the RG equation for the ghost propagator has the  IR asymptotic solution with power behavior, if the RG functions in the RG equations have the power series expansion around the IR limit of the coupling constant $g^2(\mu=0)$, see Figure~\ref{fig:RGfunct-pfunct2}. Here we use the fact that the gluon propagator has the IR asymptotic solution with power behavior, which follows from the superconvergence argument above, and that the $\beta$ function is common to both RG equations.  Thus the power solution $D_T(Q^2) \cong A(Q^2)^{\alpha-1}$ for the gluon propagator yields the power solution $\Delta_{FP}(Q^2) \cong B(Q^2)^{\beta-1}$ for the ghost propagator whose IR critical exponent is characterized by the anomalous dimension of the ghost.

Next, we must answer how to obtain the exponent $\beta$ in the IR asymptotic solutions for the ghost propagator. 
 The non-renormalization of the renormalization factor $\tilde{Z}_1$ for the gluon--ghost--antighost vertex in the Landau gauge yields the relation $\beta=-\alpha/2=-\kappa$, by supposing the existence of an IR fixed point, i.e., a finite value of $\alpha_s(0) :={g^2(0) \over 4\pi}$, suggested from the solution of the SD equation.
Thus the ghost propagator in the Landau gauge must behave like 
\begin{equation}
 \Delta_{FP}(Q^2) \cong (Q^2)^{-3/2}
\end{equation}
 in the IR region, provided that $\kappa=1/2$. 
This result shows that the color confinement criterion due to Kugo and Ojima 
\begin{equation}
 \lim_{Q^2 \rightarrow 0} [Q^2 \Delta_{FP}(Q^2)]^{-1} \equiv 1+u(0) = 0
\end{equation}
is fulfilled in the Landau gauge Yang--Mills theory.
These results should be compared with numerical simulations \cite{Bonnetetal00}. 
An exception to the above argument is the case of the IR attractor, namely the $\beta$ function has a pole. In this case, we have not yet found the consistent solution. 

In section 7, we analyze the modified MA gauge. 
The MA gauge looks quite different reflecting a fact that the propagators are gauge dependent quantities.  We show that the off-diagonal gluon and off-diagonal ghost propagators converge to $Q^2$-independent constants in the IR limit.  This implies that the off-diagonal modes become massive as  is expected from the Abelian dominance \cite{tHooft81,EI82,SY90}. 
The Abelian dominance was predicted in \cite{EI82} and subsequently  confirmed by numerical simulations on a lattice \cite{SY90}.  
On the other hand, independent analytical investigations suggest that the vacuum condensates of mass dimension 2 may be an origin of dynamical mass generation for off-diagonal modes, see \cite{Schaden99,KS00,DV03,Sawayanagi03} for the BRST non-invariant ghost condensate and  \cite{Kondo01,Kondo03b} for the BRST-invariant gluon--ghost condensate.  However, the precise relationship between the condensate and the IR asymptotic behavior is not yet fully understood. 
A remarkable result in this article is that the diagonal gluon propagator can  be of  massive type and that only the diagonal ghost propagator can be short-range, since only the diagonal ghost does not obey superconvergence. 
 The implication of this result to color confinement will be discussed in this article. 
 Our results can be compared with the recent numerical results  \cite{AS99,BCGMP03}. 

The final section is devoted to conclusion and discussion. Some of the details of the calculations are given in Appendix A.

\section{The generalized Lorentz gauge and Maximal Abelian gauge}

In this section we introduce the generalized Lorentz gauge and the modified Maximal Abelian (MA) gauge which are used in this article for fixing the gauge of the $SU(N_c)$ Yang--Mills theory.

\subsection{Generalized Lorentz gauge}

First, we modify the Lorentz gauge into more general form so that the usual Lorentz gauge is obtained as a special case.  The gauge fixing (GF) term and the associated Faddeev--Popov (FP) ghost term for the generalized Lorentz gauge (so-called the Curci--Ferrari gauge \cite{CF76}) is combined  into a compact form with two gauge-fixing parameters $\alpha$ and $\alpha'$ 
(See \cite{CF76,HL81,BT81,BT82,Ojima82,Baulieu85,DJ82,DS89,Periwal95,BSNW96,BM96,KMSI02,Gracey02}):
\begin{align}
  \mathscr{L}_{\rm GF+FP}  
  =  i \bm{\delta}_{\rm B} \bar{\bm{\delta}}_{\rm B} \left( {1 \over 2} \mathscr{A}_\mu \cdot \mathscr{A}^\mu  -{\alpha \over 2}i \mathscr{C} \cdot \bar{\mathscr{C}} \right) 
+ {\alpha' \over 2} \mathscr{B} \cdot \mathscr{B} ,
\label{GFglobal}
\end{align}
where 
$\mathscr{A}_\mu, \mathscr{B}, \mathscr{C}$ and $\bar{\mathscr{C}}$ are the non-Abelian gauge field, the Nakanishi--Lautrup (NL) auxiliary field, the Faddeev--Popov (FP) ghost and antighost fields respectively, and $\bm{\delta}_{\rm B}$ and $\bar{\bm{\delta}}_{\rm B}$ are the Becchi--Rouet--Stora--Tyutin (BRST) \cite{BRST} and anti-BRST transformations \cite{antiBRST}. 
We  use the notation 
\begin{equation}
 F \cdot G :=F^A G^A, \quad 
 F^2 :=F \cdot F , \quad 
 (F \times G)^A :=f^{ABC}F^B G^C ,
\end{equation}
with $f^{ABC}$ being the structure constants of the Lie algebra $\mathscr{G}$ of the gauge group $G$ and $A,B,C=1, \cdots, N_c^2-1$ for $G=SU(N_c)$.
We adopt the convention in which the ghost and antighost fields are Hermitian:
\begin{align}
\mathscr{C}^\dagger(x) = \mathscr{C}(x), \quad
\bar{\mathscr{C}}^\dagger(x) = \bar{\mathscr{C}}(x)  ,
\end{align}
and they are independent from each other. 
The pure Yang--Mills Lagrangian and the GF+FP term are separately invariant under the BRST and anti-BRST transformations.

\par
By performing the BRST and anti-BRST transformations explicitly, we obtain
\begin{align}
 \mathscr{L}_{\rm GF+FP} =\,& 
  {\alpha+\alpha' \over 2} \mathscr{B} \cdot \mathscr{B} 
+ \mathscr{B} \cdot \partial_\mu \mathscr{A}^\mu
  - {\alpha \over 2}  g (\mathscr{C} \times i\bar{\mathscr{C}}) \cdot \mathscr{B} 
\nonumber\\&
 + i \bar{\mathscr{C}} \cdot \partial_\mu \mathscr{D}^\mu[\mathscr{A}]\mathscr{C}
+ {\alpha \over 8}g^2 (\bar{\mathscr{C}} \times \bar{\mathscr{C}}) \cdot (\mathscr{C} \times \mathscr{C}) 
\\
=\,& {\alpha+\alpha' \over 2}\mathscr{B} \cdot \mathscr{B}
+ \mathscr{B} \cdot \partial_\mu \mathscr{A}^\mu 
   - {\alpha \over 2}  g (\mathscr{C} \times i\bar{\mathscr{C}}) \cdot \mathscr{B} 
\nonumber\\&
+ i \bar{\mathscr{C}} \cdot \partial_\mu \mathscr{D}^\mu[\mathscr{A}]\mathscr{C}
+ {\alpha \over 4}g^2 (\mathscr{C} \times i\bar{\mathscr{C}}) \cdot (\mathscr{C} \times i\bar{\mathscr{C}}) .
\end{align}
The GF+FP term includes the ghost self-interaction whose strength 
is proportional to the parameter $\alpha$.
In the limit $\alpha \rightarrow 0$, the above GF+FP term reduces to the usual form of the Lorentz gauge which is found in the most textbooks of quantum field theory, 
\begin{subequations}
\begin{align}
 \mathscr{L}_{\rm GF+FP} 
 =&
i \bm{\delta}_{\rm B} \bar{\bm{\delta}}_{\rm B} \left( {1 \over 2} \mathscr{A}_\mu \cdot \mathscr{A}^\mu   \right)  
+ {\alpha' \over 2} \mathscr{B} \cdot \mathscr{B} 
\\
=& - i \bm{\delta}_{\rm B} \left( \bar{\mathscr{C}} \cdot \partial^\mu \mathscr{A}_\mu  \right) 
+ {\alpha' \over 2} \mathscr{B} \cdot \mathscr{B} 
 \\
 =& i \bar{\mathscr{C}} \cdot \partial_\mu \mathscr{D}^\mu[\mathscr{A}]\mathscr{C}
+ \mathscr{B} \cdot \partial_\mu \mathscr{A}^\mu 
+ {\alpha' \over 2} \mathscr{B} \cdot \mathscr{B}  .
\end{align}
\end{subequations}
The Landau gauge corresponds to $\alpha'=0$ in the usual Lorentz gauge.

For later convenience, we introduce another parameterization of two gauge parameters:
\begin{equation}
  \lambda := \alpha+\alpha', \quad
  \xi := {\alpha/2 \over \alpha+\alpha'} \left( = {\alpha \over 2\lambda} \right) ,
\end{equation}
to rewrite the GF+FP term into
\begin{align}
 \mathscr{L}_{\rm GF+FP} =\,& 
  {\lambda \over 2} \mathscr{B} \cdot \mathscr{B} 
+ \mathscr{B} \cdot \partial_\mu \mathscr{A}^\mu
  - \lambda \xi  g (\mathscr{C} \times i\bar{\mathscr{C}}) \cdot \mathscr{B} 
\nonumber\\&
 + i \bar{\mathscr{C}} \cdot \partial_\mu \mathscr{D}^\mu[\mathscr{A}]\mathscr{C}
+ {\lambda \xi \over 4}g^2 (\bar{\mathscr{C}} \times \bar{\mathscr{C}}) \cdot (\mathscr{C} \times \mathscr{C}) 
\\
=\,& {\lambda \over 2}\mathscr{B} \cdot \mathscr{B}
+ \mathscr{B} \cdot \partial_\mu \mathscr{A}^\mu 
   - \lambda \xi  g (\mathscr{C} \times i\bar{\mathscr{C}}) \cdot \mathscr{B} 
\nonumber\\&
+ i \bar{\mathscr{C}} \cdot \partial_\mu \mathscr{D}^\mu[\mathscr{A}]\mathscr{C}
+ {\lambda \xi \over 2}g^2 (\mathscr{C} \times i\bar{\mathscr{C}}) \cdot (\mathscr{C} \times i\bar{\mathscr{C}}) .
\label{CF2}
\end{align}
The GF+FP term (\ref{CF2}) has the reflection symmetry: the Lagrangian is invariant under the reflection with respect to $\xi=1/2$ and the simultaneous exchange between the ghost and antighost fields, i.e., 
$\xi \leftrightarrow 1-\xi$, $\mathscr{C} \leftrightarrow \bar{\mathscr{C}}$.  Therefore, the Yang--Mills theory with $\xi=0$ GF+FP term  is equivalent to that with $\xi=1$ GF+FP term. 
The $\xi=0$ case is usually used in the textbooks of quantum field theory.  
At $\xi=1/2$, the Lagrangian is invariant under the exchange of ghost and antighost.  
For $\xi=1/2$, moreover, the GF+FP term has a hidden super symmetry
$OSp(4|2)$, see \cite{Baulieu85}. 
The Landau gauge is reached by various limits, e.g.,  $\lambda \rightarrow 0$ for $\xi=1/2$ ($\alpha'=0, \alpha \rightarrow 0$), or $\lambda \rightarrow 0$ for $\xi=0$ ($\alpha=0, \alpha' \rightarrow 0$) in the usual Lorentz gauge.  Therefore, the Landau gauge can be regarded as a common point for the three invariant subspaces specified by $\xi=0, 1/2, 1$, see Fig.~4 of \cite{KMSI02} or Appendix A for more details.

\subsection{Modified Maximal Abelian gauge}

The GF+FP term in the modified Maximal Abelian (MA) gauge \cite{KondoII} is given by 
\begin{equation}
  S_{GF+FP}' = \int d^4x i \bm{\delta}_{\rm B} \bar{\bm{\delta}}_{\rm B} \left[ 
{1 \over 2} A_\mu^a(x) A^\mu{}^a(x) - {\alpha \over 2}i C^a(x) \bar C^a(x) \right] ,
\label{GF1}
\end{equation}
where  the respective field $\Phi^A = \mathscr{A}^A, \mathscr{B}^A, \mathscr{C}^A, \bar{\mathscr{C}}^A$ is decomposed into the diagonal $\Phi^i$ and off-diagonal $\Phi^a$ components ($A=1,\cdots,N_c^2-1$ for $SU(N_c)$ gauge group),
\begin{equation}
  \Phi(x) = \Phi^A(x) T^A = \Phi^i(x)T^i + \Phi^a(x) T^a  ,
\end{equation}
and the index $i$ runs over the $N_c-1$ diagonal components corresponding to the maximal torus group $U(1)^{N_c-1}$ of the gauge group $SU(N_c)$ and the index $a$ runs over the remaining $N_c^2-N_c$ off-diagonal components for $SU(N_c)/U(1)^{N_c-1}$.
The conventional choices are $a=1,2; i=3$ for $SU(2)$ and $a=1,2,4,5,6,7; i=3,8$ for $SU(3)$.  

\par
By performing the BRST transformation explicitly, we obtain
\begin{eqnarray}
  S_{GF+FP}' &=&   \int d^4x  \Big\{ 
B^a D_\mu[a]^{ab}A^\mu{}^b+ {\alpha \over 2} B^a B^a
\nonumber\\
&&+ i \bar C^a D_\mu[a]^{ac} D^\mu[a]^{cb} C^b
- i g^2 f^{adi} f^{cbi} \bar C^a C^b A^\mu{}^c A_\mu^d 
\nonumber\\
&&+ i \bar C^a D_\mu[a]^{ac}(g f^{cdb}  A^\mu{}^d C^b)
+ i \bar C^a g  f^{abi} (D^\mu[a]^{bc}A_\mu^c) C^i 
\nonumber\\
&&+{\alpha \over 8} g^2 f^{abe}f^{cde} \bar C^a \bar C^b C^c C^d
+ {\alpha \over 4} g^2 f^{abc} f^{aid} \bar C^b \bar C^c C^i C^d
+ {\alpha \over 2} g f^{abc} i B^b C^a \bar C^c  
\nonumber\\
&&- \alpha  g f^{abi} i B^a \bar C^b C^i 
+ {\alpha \over 4} g^2 f^{abi} f^{cdi} \bar C^a \bar C^b C^c C^d \Big\} .
\label{GF3}
\end{eqnarray}
In particular, the $G=SU(2)$ case is greatly simplified as
\begin{eqnarray}
  S_{GF+FP}' &=&   \int d^4x  \Big\{ 
B^a D_\mu[a]^{ab}A^\mu{}^b+ {\alpha \over 2} B^a B^a
\nonumber\\
&&+ i \bar C^a D_\mu[a]^{ac} D^\mu[a]^{cb} C^b
- i g^2 \epsilon^{ad} \epsilon^{cb} \bar C^a C^b A^\mu{}^c A_\mu^d 
\nonumber\\
&&
+ i \bar C^a g  \epsilon^{ab} (D_\mu[a]^{bc}A_\mu^c) C^3 
\nonumber\\
&&- \alpha  g \epsilon^{ab} i B^a \bar C^b C^3 
+ {\alpha \over 4} g^2 \epsilon^{ab} \epsilon^{cd} \bar C^a \bar C^b C^c C^d \Big\}  ,
\label{GF4}
\end{eqnarray}
which reduces after integrating out the NL field $B^a$ to
\begin{eqnarray}
  S_{GF+FP}' &=&   \int d^4x  \Big\{ 
-{1 \over 2\alpha}(D_\mu[a]^{ab}A^\mu{}^b)^2  
\nonumber\\
&&+i \bar C^a D_\mu[a]^{ac} D^\mu[a]^{cb} C^b
- i g^2 \epsilon^{ad} \epsilon^{cb} \bar C^a C^b A^\mu{}^c A_\mu^d 
\nonumber\\
&&+ {\alpha \over 4} g^2 \epsilon^{ab} \epsilon^{cd} \bar C^a \bar C^b C^c C^d \Big\} .
\label{GF5}
\end{eqnarray}
\par
In order to completely fix the gauge, we need the GF+FP term also for the diagonal field.
A simple choice is the Lorentz gauge for the diagonal part:
\begin{align}
  S_{GF+FP}^{diag} =& - \int d^4x \ i \bm{\delta}_{\rm B} \left[ \bar{\mathscr{C}}^i \left\{ \partial^\mu A_\mu^i(x) + {\beta \over 2} B^i(x)  \right\} \right] 
\nonumber\\
=& B^i \partial^\mu a_\mu^i + {\beta \over 2} (B^i)^2
+i \bar{C}^i\partial^2 C^i + i \bar{C}^i \partial^\mu (g f^{ibc} A_\mu^b C^c)  .
\end{align}
There are many other choices for the MA gauge.  
A characteristic feature of the MA gauge is the explicit breaking of global color symmetry in addition to the local gauge symmetry and only the global $U(1)$ symmetry remains unbroken after fixing the residual local $U(1)$ symmetry. 
For more details of the symmetry in the most general MA gauge, see \cite{SIK01}.

\section{Analyticity, multiplicative renormalizability and RG equation}

In this section we summarize for reader's convenience the basic ingredients which are necessary in the following argument, since they are scattered in various references.

\subsection{Renormalization transformation and asymptotic behavior}

We define the renormalization transformation of the field $\Phi$ by multiplying the field by a finite positive factor $Z_\Phi$  as
\begin{align}
  \Phi(x, g'{}^2, \mu'{}^2, \lambda') = Z_\Phi^{1/2} \Phi(x, g^2, \mu^2, \lambda) 
\label{rentransf2}
\end{align}
where $g$ is the coupling constant, $\mu$ the renormalization scale, and $\lambda$ is the gauge fixing parameter. 
Such a transformation is called the equivalent transformation. 
Two sets of parameters $(g,\mu, \lambda)$ and $(g',\mu', \lambda')$ are called equivalent parameter set and are denoted as $(g,\mu, \lambda) \sim (g',\mu', \lambda')$ if the corresponding field operators are related by the renormalization transformation (\ref{rentransf2}).

We define the effective coupling or invariant charge 
\begin{align}
  \mathcal{Q}=\mathcal{Q}(k^2, g^2, \mu^2) = \mathcal{Q}\left( {k^2 \over \mu^2},g^2 \right) 
\label{defQ}
\end{align}
as a dimensionless invariant of the RG which satisfies 
\begin{align}
  \mathcal{Q}(k^2,g^2,\mu^2) = g^2  \quad \text{at} \quad k^2=\mu^2 .   
\label{iniQ}
\end{align}
Here the invariance under the RG means that the relation 
\begin{align}
  \mathcal{Q}(k^2, g'{}^2, \mu'{}^2) = \mathcal{Q}(k^2, g^2, \mu^2)
\label{equivQ}
\end{align}
holds for equivalent parameter set $(g,\mu) \sim (g',\mu')$. 
The effective coupling is usually defined in terms of time-ordered functions and an explicit example will be given later. 
We have required that the effective coupling or invariant charge should not depend on the gauge parameter $\lambda$.

By putting $k^2=\mu'{}^2$ in eq.(\ref{equivQ}), we obtain
\begin{align}
  \mathcal{Q}(\mu'{}^2, g'{}^2, \mu'{}^2) = \mathcal{Q}(\mu'{}^2, g^2, \mu^2) = g'{}^2 ,
\end{align}
where we have used eq.(\ref{iniQ}) in the last equality. 
Then the definition (\ref{defQ}) implies 
\begin{align}
 g'{}^2 = \mathcal{Q}\left( {\mu'{}^2 \over \mu^2}, g^2 \right) .
\end{align}

The propagator $D$  of the field $\Phi$ is transformed like 
\begin{align}
  D(k^2, g'{}^2, \mu'{}^2, \lambda') = Z_\Phi D(k^2, g^2, \mu^2, \lambda) .
\end{align}
Moreover, we introduce the dimensionless function, say, the dressed function (or form factor) $R$ by
\begin{align}
  R\left( {k^2 \over \mu^2}, g^2, \lambda \right) := -k^2 D(k^2, g^2, \mu^2, \lambda) .
\end{align}
Then $R$ is transformed like
\begin{align}
  R\left( {k^2 \over \mu'{}^2}, g'{}^2, \lambda' \right) = Z_\Phi R\left( {k^2 \over \mu^2}, g^2, \lambda \right) .
\label{transff2}
\end{align}
By putting $k^2=\mu'{}^2$ in this equation, we obtain
\begin{align}
  R\left( 1, g'{}^2, \lambda' \right) = Z_\Phi R\left( {\mu'{}^2 \over \mu^2}, g^2, \lambda \right) ,
\end{align}
and hence $Z_\Phi$ is written as
\begin{align}
  Z_\Phi = R^{-1}\left( {\mu'{}^2 \over \mu^2}, g^2, \lambda \right)  
 R\left( 1, g'{}^2, \lambda' \right) .
\end{align}
Thus the transformation (\ref{transff2}) of  $R$ is rewritten into 
\begin{align}
  R\left( {\mu'{}^2 \over \mu^2}, g^2, \lambda \right)   R\left( {k^2 \over \mu'{}^2}, g'{}^2, \lambda' \right) =  R\left( 1, g'{}^2, \lambda' \right) R\left( {k^2 \over \mu^2}, g^2, \lambda \right) .
\label{Rtransf}
\end{align}
Finally, the gauge parameter transforms like%
\footnote{
 In the conventional Lorentz gauge (corresponding to $\xi=0$), it is well known that $Z_\lambda=Z_\mathscr{A}$. However, this is not the case in the generalized Lorentz gauge $\xi\not=0, 1$ except for the Landau gauge $\lambda=0$.  In fact, the one-loop calculation \cite{KMSI02} in the dimensional regularization leads to 
$
 Z_\lambda=Z_\mathscr{A}+\lambda \xi(1-\xi){(g\mu^{-\epsilon})^2 \over (4\pi)^2}{C_2(G) \over \epsilon}
$ 
where $\epsilon:=2-D/2$.
}
\begin{subequations}
\begin{align}
  \lambda' = \lambda'(g'{}^2, \lambda) 
= \lambda Z_\lambda ,
\end{align}
and in particular, for $\lambda=0$, $\xi=0$ or $\xi=1$
\begin{align}
  \lambda' = \lambda'(g'{}^2, \lambda) 
= \lambda Z_\mathscr{A} 
= \lambda R^{-1}\left({\mu'{}^2 \over \mu^2}, g^2, \lambda \right)  R\left( 1, g'{}^2, \lambda' \right) .
\end{align}
\end{subequations}

 The conventional normalization condition for the propagator:
\begin{align}
  -k^2 D(k^2, g^2, \mu^2, \lambda) = 1 \quad \text{at} \quad k^2=\mu^2 < 0 
\end{align}
leads to a normalization:
\begin{align}
 R(1,g^2, \lambda) = 1 .
\end{align}
Thus, under the conventional normalization condition, the renormalization transformation of $R$ is given by  
\begin{align}
  R \left({k^2 \over \mu^2}, g^2, \lambda \right) 
= R\left({\mu'{}^2 \over \mu^2}, g^2, \lambda \right) R\left({k^2 \over \mu'{}^2},g'{}^2, \lambda' \right) ,
\quad 
 g'{}^2 = \mathcal{Q}\left( {\mu'{}^2 \over \mu^2}, g^2 \right)
\end{align}
Here the gauge parameter transforms in particular for $\lambda=0$, $\xi=0$ or $\xi=1$ like 
\begin{align}
  \lambda' = \lambda'(g'{}^2, \lambda) 
= \lambda Z_\mathscr{A} = \lambda R^{-1}\left({\mu'{}^2 \over \mu^2}, g^2, \lambda \right)  
\end{align}
with the normalization
\begin{align}
  \lambda'(g^2, \lambda)  = \lambda .
\end{align}

\subsection{Non-perturbative multiplicative renormalizability and RG equation}

We require the {\it non-perturbative multiplicative renormalizability}.  By introducing  the renormalization factor $Z_\Phi$ for the field $\Phi$ with the renormalization scale $\mu$ and the ultraviolet cutoff $\Lambda$, 
the bare (unrenormalized) $n$-point function $\tilde{\Gamma}_{0}^{(n)}$ is related to the renormalized $n$-point function $\tilde{\Gamma}^{(n)}$ as 
\begin{align}
  \tilde{\Gamma}_{0}^{(n)}(p_1, \cdots, p_n; \Lambda^2, \alpha_0, \lambda_0) 
 = Z_{\Phi}(\mu^2,\Lambda^2)^{-n/2} \tilde{\Gamma}^{(n)}(p_1, \cdots, p_n; \mu^2,\alpha,\lambda) ,
\end{align}
where $\alpha_0$ is the bare coupling and $\alpha:=g^2/4\pi$ is the renormalized coupling constant.

The differential form of the renormalization transformation for the Green functions is the  renormalization group (RG) equation. 
 The trivial relation 
\begin{align}
 {d \over d\mu}\tilde{\Gamma}_{0}^{(n)}(p_1, \cdots, p_n; \Lambda^2,\alpha_0,\lambda_0)=0  
\end{align}
implies that the renormalized $n$-point function $\tilde{\Gamma}^{(n)}$  obeys the RG equation:%
\footnote{For the generalized Lorentz gauge, we assume $\xi$ is fixed.  In fact, this is possible for $\xi=0,1/2,1$, since the relevant   Lagrangian  has distinct global symmetry and the RG flows are restricted into the subspace specified by $\xi=0,1/2,1$ respectively, see \cite{KMSI02}.
}
\begin{align}
  \left[ \mu {\partial \over \partial \mu} + \beta(\alpha) {\partial \over \partial \alpha} 
- 2 \lambda \gamma_{\lambda}(\alpha) {\partial \over \partial \lambda} 
- n \gamma_\Phi(\alpha)   \right] \tilde{\Gamma}^{(n)}(p_1, \cdots, p_n; \mu^2,\alpha, \lambda) = 0  ,
\label{RGE1}
\end{align}
where we have defined the RG functions by%
\footnote{
The renormalization relation 
$g_R=Z_g^{-1}g$
implies the mutual relationship:
\begin{align}
  {\beta(g) \over g}={1 \over g} \mu{\partial g \over \partial \mu}
= {1 \over 2g^2}  \mu{\partial g^2 \over \partial \mu}
= {\beta(\alpha) \over 2\alpha} 
=  \mu{\partial \ln g \over \partial \mu}
= -  \mu{\partial \ln Z_g \over \partial \mu} .
\end{align}
Therefore, the $\beta$ function can be calculated by differentiating $Z_g$ by $\mu$. 
\begin{align}
   \beta(g) = -  g \mu{\partial \ln Z_g \over \partial \mu} ,
\quad
  \beta(\alpha) 
= -  2\alpha \mu{\partial \ln Z_g \over \partial \mu} .
\end{align}
Here we have used the relation 
$\lambda = \lambda_0 Z_{\lambda}^{-1}$ 
to yield 
\begin{align}
   \mu  {\partial \lambda \over \partial \mu}
=   \lambda \mu  {\partial \ln \lambda \over \partial \mu}
=  \lambda \mu  {\partial \ln \lambda_0 Z_{\lambda}^{-1} \over \partial \mu}
 = - \lambda \mu  {\partial \ln Z_{\lambda} \over \partial \mu}
= -2 \lambda \gamma_{\lambda} .
\end{align}
}
\begin{align}
  \beta(\alpha) := \mu {\partial \alpha(\mu) \over \partial \mu} ,
\quad 
 \gamma_\Phi(\alpha) := {1 \over 2} \mu {\partial \ln Z_\Phi(\mu) \over \partial \mu} ,
\quad 
 \gamma_{\lambda}(\alpha) := {1 \over 2}    \mu  {\partial \ln Z_{\lambda}(\mu) \over \partial \mu} .
\end{align}

 It is shown  \cite{homoRG} that the general solution of the RG equation (\ref{RGE1}) is given by
\begin{align}
  & \tilde{\Gamma}^{(n)}(sp_1, \cdots, sp_n; \mu^2,\alpha,\lambda)
\nonumber
  \\
  &=  s^{4-n} \exp \left[ -n \int_{0}^{t} dt' \gamma_\Phi(\bar{\alpha}(t'))\right] \tilde{\Gamma}^{(n)}(p_1, \cdots, p_n; \mu^2,\bar{\alpha}(t),\bar{\lambda}(t))
\nonumber
  \\
 &= s^{4-n} \exp \left[ -n  \int_{\alpha}^{\bar{\alpha}} d\alpha' {\gamma_\Phi(\alpha') \over \beta(\alpha')}  \right] \tilde{\Gamma}^{(n)}(p_1, \cdots, p_n; \mu^2,\bar{\alpha}(t),\bar{\lambda}(t)) ,
  \\
  t &:=   \int_{\alpha}^{\bar{\alpha}(t)} {d\lambda \over \beta(\lambda)} = \ln s , 
\end{align}
where 
we have taken into account the fact that 
$\text{dim}[\tilde{\Gamma}^{(n)}]=4-n$
for the field $\Phi$ with  $\text{dim}[\Phi]=1$, and  
$\bar{\alpha}(t)$ and $\bar{\lambda}(t)$ are the running coupling constant and the running gauge parameter.

In particular, the unrenormalized (transverse) gluon and ghost form factors $F_0, G_0$ are related to the renormalized ones $F_R, G_R$ as
\begin{align}
  F_0(p^2,\Lambda^2,\alpha_0,\lambda_0) =& Z_3(\mu^2,\Lambda^2) F_R(p^2,\mu^2,\alpha,\lambda) ,
  \\
  G_0(p^2,\Lambda^2,\alpha_0,\lambda_0) =& \tilde{Z}_3(\mu^2,\Lambda^2) G_R(p^2,\mu^2,\alpha,\lambda) .
\end{align}
The renormalized form factor $F_R$  obey the RG equation:
\begin{align}
\left[ \mu {\partial \over \partial \mu} + \beta(g^2) {\partial \over \partial g^2} 
- 2 \lambda \gamma_{\lambda}(g^2) {\partial \over \partial \lambda} 
+ 2 \gamma_{\mathscr{A}}(g^2)     \right]
 F_R\left(p^2, \mu^2,g^2,\lambda \right)  = 0 .
\end{align}
The general solution is given by
\begin{align}
  F_R(sp^2,\mu^2,\alpha,\lambda)
  =  \exp \left[ 2  \int_{\alpha}^{\bar{\alpha}} d\alpha' {\gamma_\mathscr{A}(\alpha') \over \beta(\alpha')}  \right] 
 F_R(p^2; \mu^2,\bar{\alpha}(t),\bar{\lambda}(t)) .
\end{align}

In addition, 
for  the general form of the full gluon propagator given by
\begin{align}
  D_{\mu\nu}(p) = {F(p^2) \over p^2} \left( g_{\mu\nu} - {p_\mu p_\nu \over p^2} \right) +  \lambda {F^L(p^2) \over p^2}  {p_\mu p_\nu \over p^2}  ,
\label{gluonpropa}
\end{align}
the unrenormalized longitudinal gluon form factor $F_0^L$ is related to the renormalized one $F_R^L$ as
\begin{align}
  F_0^L(p^2,\Lambda^2,\alpha_0,\lambda_0) =& Z_\lambda^{-1}(\mu^2,\Lambda^2) Z_3(\mu^2,\Lambda^2) F_R^L(p^2,\mu^2,\alpha,\lambda) .
\end{align}
The renormalized longitudinal form factor $F_R^L$ obeys the RG equation:
\begin{align}
\left[ \mu {\partial \over \partial \mu} + \beta(g^2) {\partial \over \partial g^2} 
- 2 \lambda \gamma_{\lambda}(g^2) {\partial \over \partial \lambda} 
+ 2 \gamma_{\mathscr{A}}(g^2)  -  2 \gamma_{\lambda}(g^2)   \right]
 F_R^L\left(p^2, \mu^2,g^2,\lambda \right)  = 0 .
\end{align}
The general solution is given by
\begin{align}
  F_R^L(sp^2,\mu^2,\alpha,\lambda)
  = \exp \left[ 2  \int_{\alpha}^{\bar{\alpha}} d\alpha' {\gamma_\mathscr{A}(\alpha')-\gamma_\lambda(\alpha') \over \beta(\alpha')}  \right] 
 F_R^L(p^2; \mu^2,\bar{\alpha}(t),\bar{\lambda}(t)) .
\end{align}
In the conventional Lorentz gauge, $\gamma_\mathscr{A}(\alpha)=\gamma_\lambda(\alpha)$  and the longitudinal form factor of the gluon does not run, i.e., $F_R^L(p^2,\mu^2,\alpha,\lambda)$ is a $p^2$-independent constant, as is well known.   This is not the case in the generalized Lorentz gauge. 

 It is well known that the Yang--Mills theory is multiplicatively renormalizable at least in the perturbation theory.
 The RG functions can be reliably calculated by perturbation theory in the UV region by virtue of asymptotic freedom
\begin{align}
  \beta(\alpha) = -{\beta_0 \over 2\pi} \alpha^2 + \cdots, \
  \gamma_\Phi(\alpha) = - {\gamma_0^\Phi \over 4\pi} \alpha + \cdots .
\end{align}
Therefore, the Yang--Mills theory has an UV fixed point at $\alpha=0$.
The UV asymptotic form of the $n$-point function is given by 
\begin{align}
  & \tilde{\Gamma}^{(n)}(sp_1, \cdots, sp_n; \mu^2,\alpha,\lambda)
\nonumber
  \\ &
  =  s^{4-n} \left( {\bar{\alpha}(t) \over \alpha(1)} \right)^{-{n \over 2}{\gamma_0^\Phi \over \beta_0}} \tilde{\Gamma}^{(n)}(p_1, \cdots, p_n; \mu^2,\bar{\alpha}(t),\bar{\lambda}(t)) .
\end{align}
Thus the UV behavior of the propagator $(p^2 \rightarrow \infty)$ is given by 
\begin{align}
  G_\Phi^{(2)}(p) \sim {1 \over p^{2}}  \left( {\bar{\alpha}(p) \over \alpha(\mu)} \right)^{{\gamma_0^\Phi \over \beta_0}} , 
\end{align}
for the running coupling constant
\begin{align}
   \bar{\alpha}(p) \sim \left( {\beta_0 \over 4\pi} \ln {p^2 \over \Lambda_{QCD}^2} \right)^{-1} .
\end{align}

\subsection{Renormalization group on the complex plane}

The equation (\ref{Rtransf}) or (\ref{RGE1}) is valid also for complex $k^2$.  Then the functions $D$ and $R$ satisfy the same RG equation:
\begin{align}
\left[ \mu {\partial \over \partial \mu} + \beta(g^2) {\partial \over \partial g^2} 
- 2 \lambda \gamma_{\lambda}(g^2) {\partial \over \partial \lambda} 
+ 2 \gamma_{\mathscr{A}}(g^2)   \right]
 R \left({k^2 \over \mu^2},g^2,\lambda \right)  = 0 ,
\label{RGE}
\end{align}
with the same coefficients, i.e., RG functions as those of (\ref{RGE1}) at every point $k^2$ of the cut $k^2$ plane (i.e.,  complex $k^2$ plane except for the positive real axis).


\begin{figure}[htbp]
\begin{center}
\includegraphics[height=8cm]{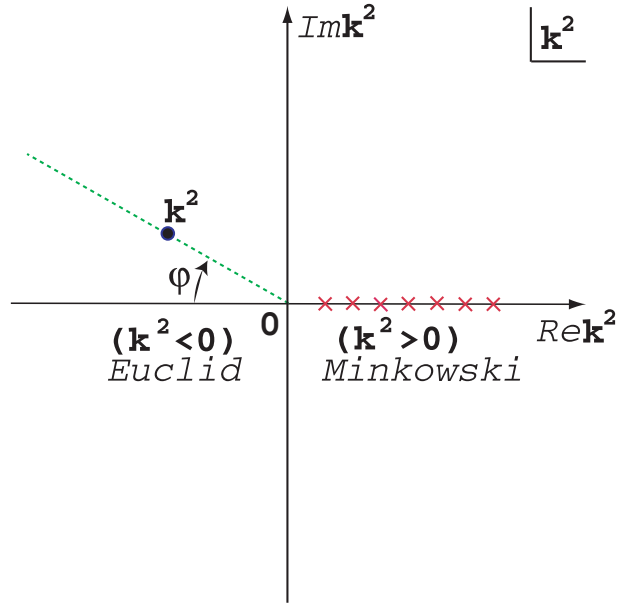}
\caption{The ray with angle $\varphi$ measured from the negative real axis $k^2<0$ on the complex $k^2$ plane with singularities on the positive real axis.}
\label{fig:ray}
\end{center}
\end{figure}


We consider the propagator along the ray in the cut $k^2$ plane (See Figure~\ref{fig:ray}):
\begin{align}
  k^2 = - |k^2| e^{i\varphi} = |k^2| e^{i(\pi+\varphi)} 
, \quad - \pi < \varphi  < \pi ,
\label{ray2}
\end{align}
where the angle $\varphi$ is measured from the negative real axis $k^2<0$ in the direction indicated by the arrow in Figure~\ref{fig:ray}. 
Setting $\mu'{}^2=-|k^2|$, the relation (\ref{Rtransf}) yields
\begin{align}
  R \left({k^2 \over \mu^2}, g^2, \lambda \right) 
= R\left({-|k^2| \over \mu^2}, g^2, \lambda  \right) R\left({k^2 \over -|k^2|}, \bar{g}^2, \bar{\lambda} \right) 
R^{-1}\left(1, \bar{g}^2, \bar{\lambda} \right) 
,
\quad 
 \bar{g}^2 = \mathcal{Q}\left( {-|k^2| \over \mu^2}, g^2 \right) ,
\end{align}
which is rewritten as
\begin{align}
  R \left({k^2 \over \mu^2}, g^2, \lambda  \right) 
= R\left({|k^2| \over |\mu^2|}, g^2, \lambda  \right) R\left(e^{i\varphi} ,Q, \bar{\lambda}  \right)
R^{-1}\left(1, \bar{g}^2, \bar{\lambda} \right)
 ,
\quad
 \bar{g}^2 = \mathcal{Q} = \mathcal{Q}\left( {|k^2| \over |\mu^2|}, g^2 \right) .
\end{align}
In addition, the gauge parameter is transformed in particular  for $\lambda=0$, $\xi=0$ or $\xi=1$ as
\begin{align}
  \bar{\lambda} = \lambda R^{-1}\left({|k^2| \over |\mu^2|}, g^2, \lambda \right)  R\left(1, \bar{g}^2, \bar{\lambda} \right) .
\end{align}

Using the asymptotic behavior of the structure function for $g^2 \rightarrow +0$
\begin{align}
  R(k^2, g^2, \mu^2, \lambda) = R \left({k^2 \over \mu^2}, g^2, \lambda \right) \cong 1 + \gamma^\Phi_0(\lambda) g^2 \ln {k^2 \over \mu^2} + O(g^4) ,
\end{align}
and the continuation, 
\begin{align}
  R\left(e^{i\varphi} ,\mathcal{Q}, \bar{\lambda} \right) \cong 1 + \gamma^\Phi_0(\bar{\lambda}) \bar{g}^2 i \varphi + O(\bar{g}^4) ,
\label{continue2}
\end{align}
we obtain
\begin{align}
  \lim_{|k^2| \rightarrow \infty} R\left(e^{i\varphi} ,\mathcal{Q}, \bar{\lambda} \right) = \lim_{\mathcal{Q} \rightarrow 0} R\left(e^{i\varphi} ,\mathcal{Q}, \bar{\lambda} \right) = 1 .
\end{align}

Now, it is important to recall the asymptotic formula in Euclidean region $k^2<0$ which is obtained from the general solution of the RG equation (\ref{RGE}):
\begin{align}
  R\left( {k^2 \over \mu^2}, g^2, \lambda \right) &=  R(1,\mathcal{Q}, \bar{\lambda}) \exp \left[ 2 \int_{g^2}^{\mathcal{Q}} dx {\gamma_\Phi(x) \over \beta(x)}  \right] 
 \cong C_\Phi(\lambda) \left( \ln { k^2 \over \mu^2 } \right)^{-\gamma^\Phi_0(\lambda)/\beta_0} ,
\\
 C_\Phi(\lambda) &=  (g^2|\beta_0|)^{-\gamma^\Phi_0(\lambda)/\beta_0} \exp \left[ \int_{g^2}^{0} dx \tau_\Phi(x) \right] > 0 .
\end{align}
In the conventional normalization $R(1,\mathcal{Q},\bar{\lambda})=1$.  In other cases, the leading behavior of the coefficient $R(1,\mathcal{Q},\bar{\lambda})$ for $g^2 \rightarrow +0$ or $u:=k^2/\mu^2 \rightarrow \infty$ is determined by the asymptotic form of $R(1,g^2,\lambda)$ for small $g$.  The behavior of the exponential factor follows from the behavior of $\mathcal{Q}$ and the ratio $\gamma_\Phi/\beta$.  
Here the integrand $\gamma_\Phi(x)/\beta(x)$ is not integrable at $x=0$ in QCD and therefore the singular part is separated
$\gamma_\Phi(x)/\beta(x)=\gamma^\Phi_0/\beta_0 x^{-1} + \tau_\Phi(x)$
with $\tau_\Phi(x)$ being the remainder integrable at $x=0$. 
The coefficient $C_\Phi$ is positive.  See Appendix A. 

Thus we find the asymptotic form for $|k^2| \rightarrow \infty$ along the ray (\ref{ray2}):
\begin{align}
 R(k^2, g^2, \mu^2, \lambda) \cong & R_{as}(k^2,g^2,\mu^2, \lambda) = 
 C_\Phi(\lambda) \left( \ln {|k^2| \over |\mu^2|} \right)^{-\gamma^\Phi_0(\lambda)/\beta_0} ,
\\
 D(k^2, g^2, \mu^2, \lambda) \cong & D_{as}(k^2,g^2,\mu^2, \lambda) = 
- C_\Phi(\lambda) k^{-2} \left( \ln {|k^2| \over |\mu^2|} \right)^{-\gamma^\Phi_0(\lambda)/\beta_0} ,
\\
 C_\Phi(\lambda) =& (g^2|\beta_0|)^{-\gamma^\Phi_0(\lambda)/\beta_0} \exp \left[ \int_{g^2}^{0} dx \tau(x) \right] > 0 .
\label{alldirection2}
\end{align}

\subsection{Analyticity}


\begin{figure}[htbp]
\begin{center}
\includegraphics[height=8cm]{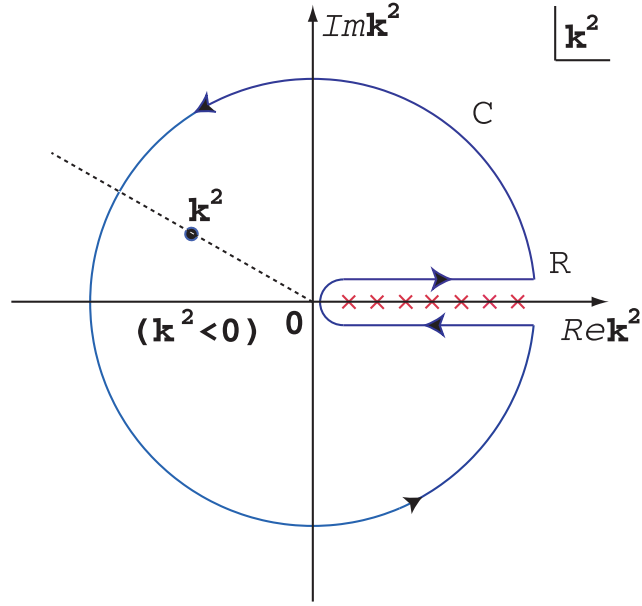}
\caption{An integration contour $C$ on the complex $k^2$ plane with singularities on the positive real axis.}
\label{fig:CutComplexPlane}
\end{center}
\end{figure}


If the analytic function $f(z)$ vanishes along any ray in the cut complex plane as $|z| \rightarrow \infty$, then it has the dispersion relation as follows. 

Suppose that the complex function $f(z)$ is analytic in the whole complex $z:=k^2$ plane except for the positive real axis $z > 0$. 
Then we choose a closed loop $C$ in the complex $k^2$ plane such that the complex function $f(z)$ is analytic inside and on the closed loop $C$ as in Figure~\ref{fig:CutComplexPlane}.
For a reference point $k^2$ inside $C$, the Cauchy integral formula tells us that for the closed contour $C$ of integration
\begin{align}
  f(k^2) =& {1 \over 2\pi i} \oint_{C} dz {f(z) \over z-k^2} 
\\
  =&    {1 \over 2\pi i} \int_{0}^{R} dz {f(z+i\epsilon)-f(z-i\epsilon) \over z-k^2} 
+ {1 \over 2\pi i} \oint_{|z|=R} dz {f(z) \over z-k^2} ,
\label{Cif}
\end{align}
where we have separated the integral into two pieces; one is the contribution from the paths above and below the positive real axis; the other is the contribution to the integral over the circle of radius $R$.  

First, making use of the fact that $f(k^2)$ is real for $k^2<s_{min}$ (at least for the space-like region $k^2<0$), we can use the Schwartz reflection principle 
$f(z-i\epsilon)=f^{*}(z+i\epsilon)$  for $Re z>s_{min}$ and hence
$f(z+i\epsilon)-f(z-i\epsilon)=f(z+i\epsilon)-f^{*}(z+i\epsilon)
= 2i Im f(z+i\epsilon)$.
Thus the first term in RHS of (\ref{Cif})  picks up the discontinuity  (the imaginary part) of the $f(z)$ function along the positive real axis.  
Next, the circle $|z|=R$ is parameterized as $z=Re^{i\theta}$:
\begin{align}
  f(k^2) =    {1 \over \pi} \int_{0}^{R} dz {Im f(z+i\epsilon) \over z-k^2} 
+  \int_{0}^{2\pi} {d\theta \over 2\pi}{Re^{i\theta} f(Re^{i\theta}) \over Re^{i\theta}-k^2} .
\end{align}
The second term in RHS of this equation has the upper bound
\begin{align}
  \Big| \int_{0}^{2\pi} {d\theta \over 2\pi} {Re^{i\theta} f(Re^{i\theta}) \over Re^{i\theta}-k^2} \Big|
  \le \int_{0}^{2\pi} {d\theta \over 2\pi}{R|f(Re^{i\theta})| \over |Re^{i\theta}- k^2|}
  \le   {\rm max}_{0<\theta<2\pi}|f(Re^{i\theta})| 
{R \over R-|k^2|} .
\end{align}
Now we wish to put the radius $R$ of the circle to infinity. 
Thus, if $|f(z)| \rightarrow 0$ as $|z| \rightarrow \infty$, then the second term has no contribution as $R \rightarrow \infty$ and we have the dispersion relation:%
\footnote{
In general, the boundary value of the analytic function is defined as the Sato hyperfunction, which is an extension of the Schwartz distribution. 
In applying this relation to the propagator and the form factor later, we require the positivity in Euclidean region, i.e, on the negative real axis of the complex $k^2$ plane.  Then they become a positive measure and can not be a distribution.  
}
\begin{align}
  f(k^2) =  {1 \over \pi} \int_{0}^{\infty} dz {Im f(z+i\epsilon) \over z-k^2} .
\label{dispersionrel0}
\end{align}
This is also written as
\begin{align}
  f(k^2) =   \int_{0}^{\infty} ds {\rho(s) \over s-k^2} , 
  \quad
  \rho(s) := {1 \over \pi} Im f(s+i0) ,
\label{dispersionrel}
\end{align}
where $k^2$ is an arbitrary point in the complex plane except for the positive real axis. 
Note that $f(z)$ need not to go to zero in the power-like, i.e., $f(z) \sim 1/|z|^{\varepsilon} (\varepsilon>0)$ as is usually assumed. 
In fact, a weaker decay $f(z) \sim (\log|z|)^{-\varepsilon} (\varepsilon>0)$ is also allowed.
Incidentally, since the Cauchy integral formula is generalized as
\begin{align}
  f^{(n)}(k^2) =& {n! \over 2\pi i} \oint_{C} dz {f(z) \over (z-k^2)^{n+1}} ,
\end{align}
the similar argument as above leads to
\begin{align}
  f^{(n)}(k^2) =  n! \int_{0}^{\infty} ds {\rho(s) \over (s-k^2)^{n+1}}   .
\end{align}

Note that the dispersion relations (\ref{dispersionrel0}) or (\ref{dispersionrel}) is derived from 
\begin{enumerate}
\item[i)] the analyticity of $f(z)$ except for the positive real axis,  
\item[ii)] the asymptotic vanishing of $f(z)$ as $|z| \rightarrow \infty$, 
\item[iii)] the real-valuedness of $f(z)$ on the real axis $z<s_{min}$ (at least for the negative real axis). 
\end{enumerate}
In the next section, we equate the dispersion relation with the spectral representation to extract the information on the spectrum using the asymptotic behavior of the propagator and the dressed functions. 

\subsection{Minkowski region}

In general, the propagators are tempered distributions in the Minkowski region $k^2>0$.  However, if distribution-type singularities are present only below a finite momentum value $K$, the propagators may be treated as ordinary functions for $k^2>K^2$ \cite{OZ80a}.  
In this case, therefore, the asymptotic formulas derived above  are also applied to the large Minkowski momenta with $\varphi=\pm\pi$.

If the distribution-type singularities occur for arbitrarily large momenta, the asymptotic behavior of the propagators should be studied in terms of the average values which are obtained by smearing with suitable test functions.  
It is shown \cite{OZ80a} that the averaged gluon propagator in the Minkowski region has again the same asymptotic form as given above, independent of the chosen test function. 

The asymptotic behavior for the weight function  of the spectral representation is derived as follows.
  The  weight function has the expression:
\begin{align}
  \pi \rho(k^2) = {\rm Im} D(k^2+i\epsilon) = -k^{-2} {\rm Im} R(k^2+i\epsilon) , \quad k^2 > 0 ,
\end{align}
where we have used the result of the manipulation:  
$
 {\rm Im}[R(k^2+i\epsilon)]
= -{\rm Im}[(k^2+i\epsilon)D(k^2+i\epsilon)]
=-k^2 {\rm Im}D(k^2+i\epsilon)- \epsilon {\rm Re}D(k^2+i\epsilon) 
$.
Setting $\mu'{}^2=-k^2$ in (\ref{Rtransf}), we have 
\begin{align}
  R\left( {k^2 \over \mu^2}, g^2, \lambda \right) 
= R\left( {k^2 \over |\mu^2|}, g^2, \lambda\right)   R\left( -1, \bar{g}^2, \bar{\lambda} \right)  R^{-1}\left( 1, \bar{g}^2, \bar{\lambda} \right)   ,
\quad 
 \bar{g}^2 = \mathcal{Q}\left( {k^2 \over |\mu^2|}, g^2 \right) ,
\end{align}
which leads to the imaginary part for $\mu^2<0$
\begin{align}
  \pi \rho(k^2) = -k^{-2} {\rm Im} R(k^2+i\epsilon) 
= -k^{-2} R\left( {k^2 \over |\mu^2|}, g^2, \lambda \right)  {\rm Im}  R\left( -1, \bar{g}^2, \bar{\lambda} \right)  R^{-1}\left( 1, \bar{g}^2, \bar{\lambda} \right)   ,
\end{align}
where $R\left( {k^2 \over |\mu^2|}, g^2, \lambda \right)$ and $R\left( 1, \bar{g}^2, \bar{\lambda} \right)$ are real for $k^2>0$. 
From (\ref{continue2}), we obtain
\begin{align}
   {\rm Im}  R\left( -1, \mathcal{Q}, \bar{\lambda} \right)  = \pi \gamma^\Phi_0(\bar{\lambda}) \mathcal{Q} + \mathcal{Q}^2 h  ,
\end{align}
with the remainder $h$ vanishing for $\mathcal{Q} \rightarrow 0$. 
It is shown by one-loop resumed perturbation theory that 
\begin{align}
  \mathcal{Q} \cong 4\pi |\beta_0^{-1}| \left( \ln {k^2 \over |\mu^2|} \right)^{-1} .
\end{align}
Thus we obtain the asymptotic form
\begin{align}
  \rho(k^2, g^2, \mu^2, \lambda) \cong& \rho_{as}(k^2, g^2, \mu^2, \lambda) 
\quad \text{as} \quad k^2 \rightarrow \infty ,
\\
  \rho_{as}(k^2, g^2, \mu^2, \lambda) =& -(k^2)^{-1} \gamma^\Phi_0(\lambda) \mathcal{Q} R_{as}\left( {k^2 \over |\mu^2|}, g^2, \lambda \right)
\\
=& -4\pi (k^2)^{-1} \gamma^\Phi_0(\lambda) |\beta_0^{-1}| C_\Phi(\lambda) \left( \ln { k^2 \over |\mu^2| } \right)^{-\gamma^\Phi_0(\lambda)/\beta_0-1} ,
\label{asymptoticrhoM}
\end{align}
where it is important to remember that the anomalous dimension $\gamma_\Phi$ is gauge  dependent and so is the asymptotic form.  

If the weight function has distribution-type singularities at arbitrarily large momenta, we consider the asymptotic behavior of the average values. Consequently, the same asymptotic behavior is obtained \cite{OZ80a}.  

It should be remarked that {\it the sign of the asymptotic discontinuity $\rho_{as}$ is determined by the ratio $\gamma_0/\beta_0$, i.e., the sign of $\rho$ is the opposite of $\gamma_0/\beta_0$. 
It is negative for $\gamma_0/\beta_0>0$.}

\section{Naive derivation of superconvergence relation}

Now we can derive the superconvergence relations in the generalized Lorentz gauge.   
Our results reproduce the well-known superconvergence relation in the conventional Lorentz gauge. 
 The details of calculations are given in Appendix A. 

\subsection{Gluon}

 For $\gamma^A_0/\beta_0>0$, the {\it renormalized} gluon structure function $D_T$ has the property that the complex function defined by 
\begin{align}
  D_T(k^2, \mu^2, g_R^2, \lambda_R)  + {\lambda/\lambda_* \over k^2} 
\label{Dgeneral}
\end{align}
vanishes faster than $(k^2)^{-1}$, indeed it behaves like 
$
 (k^2)^{-1} C_A \left( \ln { |k^2| \over |\mu^2| } \right)^{-\gamma^A_0/\beta_0}
$ 
for large $|k^2|$ in all directions of the complex $k^2$ plane. 
For $\gamma^A_0/\beta_0>0$, moreover, the {\it renormalized} dressed function $R=-k^2 D_T$ has the property that  the complex function defined by 
\begin{align}
  R(k^2, \mu^2, g_R^2, \lambda_R)  - {\lambda \over \lambda_*} 
\label{Rgeneral}
\end{align}
vanishes like 
$
 C_A \left( \ln { |k^2| \over |\mu^2| } \right)^{-\gamma^A_0/\beta_0} 
\rightarrow 0
$
as $|k^2| \rightarrow \infty$, 
in all directions of the complex $k^2$ plane.
 By applying the formula (\ref{dispersionrel0}) to the function (\ref{Dgeneral}), therefore, we can write an unsubtracted dispersion relation
\begin{align}
  D_T(k^2, \mu^2, g_R^2, \lambda_R) 
 = \int_{0}^{\infty} dp^2 {\rho(p^2, \mu^2, g_R^2, \lambda_R) \over p^2 -k^2} ,
\end{align}
where we have used a mathematical identity:
\begin{align}
 {1 \over p^2+i\epsilon} = P{1 \over p^2} - i\pi \delta(p^2) .
\end{align}
In the similar way,  the formula (\ref{dispersionrel0}) applied to the function (\ref{Rgeneral}) yields
\begin{align}
  R(k^2, \mu^2, g_R^2, \lambda_R) = {\lambda \over \lambda_*} - \int_{0}^{\infty} dp^2  {p^2 \rho(p^2, \mu^2, g_R^2, \lambda_R) \over p^2 -k^2} .
\end{align}
Therefore there are two alternative forms for the spectral representation for $D$:
\begin{align}
  D_T(k^2, \mu^2, g_R^2, \lambda_R) = \int_{0}^{\infty} dp^2 {\rho(p^2, \mu^2, g_R^2, \lambda_R) \over p^2 -k^2} 
= - {\lambda/\lambda_* \over k^2}
+ {1 \over k^2} \int_{0}^{\infty} dp^2  {p^2 \rho(p^2, \mu^2, g_R^2, \lambda_R) \over p^2 -k^2} .
\end{align}
This is possible if and only if the relation 
\begin{align}
  0 =  {1 \over k^2} \left[ \int_{0}^{\infty} dp^2  \rho(p^2, \mu^2, g_R^2, \lambda_R) - {\lambda \over \lambda_*}  \right] 
\end{align}
is satisfied for arbitrary $k^2$. 
 For $\gamma^A_0/\beta_0>0$, thus, we have the superconvergence relation in the generalized Lorentz gauge with an {\it initial} gauge-fixing parameter $\lambda$:
\begin{align}
  \int_{0}^{\infty} dk^2 \rho(k^2, \mu^2, g_R, \lambda_R) = {\lambda \over \lambda_*} .
\label{gluonsuperconv}
\end{align}
This relation reduces to the well-known superconvergence relation in the Landau gauge
\begin{align}
  \int_{0}^{\infty} dk^2 \rho(k^2, \mu^2, g_R, 0) = 0 .
\end{align}
Here $\lambda_*$ depends on $\xi$ in the generalized Lorentz gauge:  $\lambda_*=13/3$ in the conventional Lorentz gauge $\xi=0$ and
$
 \lambda_* = {13 \over 3}{1 \over 1-\xi} \ \text{for} \ \xi<1/2 , 
$
$ 
= {13 \over 3}{1 \over \xi}   \ \text{for} \ \xi>1/2 
$,
see section V of \cite{KMSI02} and Appendix A.

Irrespective of the gauge parameter, therefore, the gluon form factor vanishes in the IR limit due to superconvergence relation (\ref{gluonsuperconv}):
\begin{align}
  R(0, \mu^2, g_R^2, \lambda_R) = {\lambda \over \lambda_*} - \int_{-0}^{\infty} dp^2 \rho(p^2, \mu^2, g_R^2, \lambda_R)  
= 0 ,
\end{align}
and hence the gluon propagator and the form factor has the gauge-parameter independent IR behavior for the small momenta: 
\begin{align}
 F(Q^2, \mu^2, g_R^2, \lambda_R) = O(Q^2) ,
\quad D(Q^2, \mu^2, g_R^2, \lambda_R) = \text{const.} + O(Q^2) .
\end{align}
These behavior are independent from the gauge parameter. 

\subsection{Ghost}

However, the situation is quite different for the ghost.  
We can repeat the same steps as in the gluon case. Then we obtain the following results which agree with those of 
\cite{Xu96} obtained in the conventional Landau gauge. 
In the arbitrary gauge, the ghost has an unsubtracted dispersion relation  for the {\it renormalized} ghost propagator $\Delta_{FP}$: 
\begin{align}
  \Delta_{FP}(k^2, \mu^2, g_R^2, 0) 
 =& \int_{0}^{\infty} dp^2 {\rho_{FP}(p^2, \mu^2, g_R^2, 0) \over p^2 -k^2} .
\end{align}
However, 
the form factor $G_{FP}:=-k^2 \Delta_{FP}$ has an unsubtracted dispersion relation  
only in the Landau gauge $\lambda=0$.
\begin{align}
  G_{FP}(k^2, \mu^2, g_R^2, 0) =&   - \int_{-0}^{\infty} dp^2  {p^2 \rho_{FP}(p^2, \mu^2, g_R^2, 0) \over p^2 -k^2} .
\end{align}
Hence, the superconvergence relation for the ghost holds only in the Landau gauge $\lambda=0$
\begin{align}
  0 = & \int_{0}^{\infty} dk^2 \rho_{FP}(k^2, \mu^2, g_R^2, 0) .
\end{align}
For $\lambda\not=0$, however, the superconvergence relation does not hold, since  the unsubtracted dispersion relation exists only for the propagator, not for the form factor:
\begin{align}
  \Delta_{FP}(k^2, \mu^2, g_R^2, \lambda_R) 
 = \int_{0}^{\infty} dp^2 {\rho_{FP}(p^2, \mu^2, g_R^2, \lambda_R) \over p^2 -k^2} 
= -{1 \over k^2} G_{FP}(k^2, \mu^2, g_R^2, \lambda_R) .
\end{align}
If this result is true, then the superconvergence for the ghost holds only in the Landau gauge, in sharp contrast with the gluon case. 
There is a subtle point which may invalidate the above derivation.  We reconsider the ghost in the next section.

\section{Superconvergence relation and spectrum}

In this section, we combine the superconvergence relation with the spectral representation to discuss the spectrum of the Yang--Mills theory.

\subsection{Spectral representation}

 From the assumptions:
1) Poincar\'e group structure (representation),
2) spectral condition,
3) completeness condition,
the spectral representation (Umezawa--Kamefuchi--K\"allen--Lehmann representation) \cite{spectralrep} for the gluon two-point function  follows:
\begin{align}
 D_{T}(k^2)  
= \int_{0}^{\infty} d\sigma^2 \rho(\sigma^2) {1 \over \sigma^2-k^2-i \epsilon}  ,
\label{specrep}
\end{align}
where $\rho(\sigma^2)$ is the spectral function. 
 Due to a mathematical identity 
\begin{align}
 {1 \over x-i\epsilon} = P{1 \over x} + i\pi \delta(x) ,
\end{align}
 the spectral representation (\ref{specrep}) of the  propagator  
$
  D_{T}(k^2) = Re D_{T}(k^2) + i Im D_{T}(k^2)
$
is decomposed into two parts:
\begin{align}
 Im D_{T}(k^2)  = \pi  \rho(k^2) ,
\end{align}
and
\begin{align}
 Re D_{T}(k^2)  = \int_{0}^{\infty} d\sigma^2 \rho(\sigma^2) P{1 \over \sigma^2-k^2} .
\end{align}
Thus we obtain the same relation as the dispersion relation:
\begin{align}
 Re D_{T}(k^2)  = P \int_{0}^{\infty} d\sigma^2 \rho(\sigma^2)  {1 \over \sigma^2-k^2} , \quad
\rho(p^2) = {1 \over \pi} Im D_{T}(p^2) .
\end{align}
However, it should be remarked that we have not used the analyticity for deriving this representation.

If the state space has positive definite inner product, the spectral function must be non-negative: 
\begin{equation}
  \rho(p^2) \ge 0 .
\end{equation}
In the state space with the indefinite inner product, the complete set $\{ |n \rangle \}$ can not be spanned by the eigenstates of $P^\mu$ alone in general and the metric 
\begin{equation}
  \langle n | n' \rangle = \eta_{nn'} 
\end{equation}
can not be diagonalized, and 
the completeness condition must be replaced by
\begin{equation}
   1 = \sum_{n,n'}  | n \rangle \eta_{nn'}^{-1} \langle n' | .
\end{equation}
In Yang--Mills theory, fortunately, all $|n \rangle$ can be taken to be   eigenstates of $P^\mu$ and the metric can be diagonalized into the form $\eta_{nn'} = \pm 1$.  
In this case, the positivity of the spectral function does not hold,   since both positive and negative terms contribute to the sum defining the spectral function.
This fact plays the important role in the superconvergence relation in Yang--Mills theory.

\subsection{IR behavior of the renormalized gluon propagator in Euclidean region}

The spectral representation of the renormalized gluon structure function reads
\begin{align}
  D_T(k^2, \mu^2, g_R^2,\lambda_R) =& \int_{0}^{\infty} dp^2 {\rho(p^2,\mu^2,g_R^2,\lambda_R) \over p^2-k^2}, 
\\
  \rho(p^2,\mu^2,g_R^2,\lambda_R) =& \pi^{-1} Im D_T(k^2+i\epsilon, \mu^2, g_R^2,\lambda_R) .
\label{gluonspec}
\end{align}
The renormalized version of $\rho$ is a function of $p^2, \mu, g_R, \lambda_R$ where $\mu$ is the renormalization point.
Note that this representation holds for arbitrary point $k^2$ in the complex $k^2$ plane except for the positive real axis $k^2>0$.  
It was shown that the gluon structure function has the  superconvergence relation:
\begin{align}
  \int_{0}^{\infty} dp^2 \rho(p^2,\mu^2,g_R^2,\lambda_R) 
 = {\lambda \over \lambda_*}  ,
\label{gluonsconv}
\end{align}
in the parameter region satisfying $\gamma^{A}_{0}/\beta_{0}>0$. 
In this paper we consider only the case $\beta_{0}<0$ in pure Yang-Mills theory corresponding to asymptotic freedom.
Defining the dimensionless dressed function $R$ by
\begin{align}
  R(k^2) := -k^2 D_T(k^2) ,
\end{align}
we can write the dispersion and spectral representation of the dressed function:
\begin{align}
  R(k^2, \mu^2, g_R^2,\lambda_R) = {\lambda \over \lambda_{*}} - \int_{0}^{\infty} dp^2 {p^2 \rho(p^2,\mu^2,g_R^2,\lambda_R) \over p^2-k^2}  .
\label{gluonformspec}
\end{align}

Putting the reference point $k^2$ on the Euclidean region, i.e., $Q^2=-k^2>0$, the gluon structure function has the representation:
\begin{align}
  D_T(-Q^2,\mu^2,g_R^2,\lambda_R) 
=  \int_{0}^{\infty} dp^2 {\rho(p^2,\mu^2,g_R^2,\lambda_R) \over p^2+Q^2} .
\label{EuclidD}
\end{align}
In fact, $D_T$ is not singular in the Euclidean region $Q^2>0$.
Defining the dimensionless gluon form factor $F$ by
\begin{align}
  F(Q^2) := R(-Q^2) = Q^2 D_T(-Q^2) ,
\end{align}
we can write the dispersion and spectral representation of the form factor in the Euclidean region $Q^2=-k^2>0$:
\begin{align}
  F(Q^2,\mu^2,g_R^2,\lambda_R)  
= {\lambda \over \lambda_*} - \int_{0}^{\infty} dp^2 {p^2 \rho(p^2,\mu^2,g_R^2,\lambda_R) \over p^2+Q^2} .
\label{EuclidF}
\end{align}

Confinement is considered to be a strong coupling problem and to be  related to the IR region.
Therefore, we focus on the IR behavior of the gluon form factor.  
We find that {\it the gluon form factor vanishes in the IR limit of the Euclidean region} $Q^2 \rightarrow +0$:
\begin{align}
  F(Q^2=0, \mu^2, g_R^2,\lambda_R) 
 = {\lambda \over \lambda_*} - \int_{0}^{\infty} dp^2  \rho(p^2,\mu^2,g_R^2,\lambda_R) 
= 0 ,
\end{align}
due to the superconvergence relation (\ref{gluonsconv}).
Moreover, the spectral representation has the formal power series for small $Q^2$, 
\begin{align}
  F(Q^2,\mu^2,g_R^2,\lambda_R) 
=& \sum_{n=1}^{\infty} (-1)^{n+1} (Q^2)^n 
\int_{0}^{\infty} dp^2 {\rho(p^2,\mu^2,g_R^2,\lambda_R) \over (p^2)^n}  
\nonumber\\
=&  Q^2 \int_{0}^{\infty} dp^2 {\rho(p^2,\mu^2,g_R^2,\lambda_R) \over p^2} 
- (Q^2)^2 \int_{0}^{\infty} dp^2 {\rho(p^2,\mu^2,g_R^2,\lambda_R) \over (p^2)^{2}} 
+ O(Q^6) ,
\label{Fexpan}
\end{align}
while the gluon structure function has the IR behavior
\begin{align}
  D_T(-Q^2,\mu^2,g_R^2,\lambda_R) 
=& \sum_{n=0}^{\infty} (-1)^{n} (Q^2)^{n} 
\int_{0}^{\infty} dp^2 {\rho(p^2,\mu^2,g_R^2,\lambda_R) \over (p^2)^{n+1}} 
\nonumber\\
=&  \int_{0}^{\infty} dp^2 {\rho(p^2,\mu^2,g_R^2,\lambda_R) \over p^2} 
- Q^2 \int_{0}^{\infty} dp^2 {\rho(p^2,\mu^2,g_R^2,\lambda_R) \over (p^2)^{2}} 
+ O(Q^4) .
\label{Dexpan}
\end{align}
In the Euclidean IR limit $Q^2 \downarrow 0$, the gluon structure function converges to a constant:
\begin{align}
  D_T(0,\mu^2,g_R^2,\lambda_R) 
=  \int_{0}^{\infty} dp^2 {\rho(p^2,\mu^2,g_R^2,\lambda_R) \over p^2} ,
\label{D(0)}
\end{align}
provided that all the integrals are convergent.
If this constant is non-vanishing and finite, we have 
  $\kappa=1/2$.

It is expected that each integral appearing in the coefficient of the power series (\ref{Fexpan}) and (\ref{Dexpan}) is positive and the series is alternating. 
In particular, the constant $D_T(0)$ (\ref{D(0)}) is considered to be positive.  
The integral is convergent for large $p^2$ due to the asymptotic form (\ref{asymptoticrhoM}).  For large $p^2$, the integrand $\rho/p^2$ is small and negative, while it is large positive for small $p^2$ and it does not diverge due to the existence of mass gap.  Therefore, the total integration is expected to be positive. 
This conclusion is a gauge independent result, i.e., independent of the choice of gauge parameter. 
\footnote{
Now we consider the implication of the above result to the spectrum. 
If the gluon spectral function has a massless pole and 
the single particle contribution is explicitly separated as 
$\rho(p^2) = Z \delta(p^2) + \cdots$, then $D_T(0)$ given by (\ref{D(0)}) diverges. 
Therefore,  finiteness of the renormalized gluon propagator at $Q^2=0$ excludes the massless gluon pole, suggesting the existence of mass gap. 
In this case,  all the coefficients of the power series in $Q^2$ are finite due to the asymptotic form (\ref{asymptoticrhoM}) of $\rho$ for large $p^2$.
Consequently, the IR power series expansions (\ref{Fexpan}) and (\ref{Dexpan}) are well-defined. 
}

If the constant (\ref{D(0)}) happens to vanish, 
\begin{align}
   \int_{0}^{\infty} dp^2 {\rho(p^2,\mu^2,g_R^2,\lambda_R) \over p^2} = 0 ,
\end{align}
then the $\kappa=1$ case occurs, 
\begin{align}
  F(Q^2,\mu^2,g_R^2,\lambda_R) 
= - (Q^2)^2 \int_{0}^{\infty} dp^2 {\rho(p^2,\mu^2,g_R^2,\lambda_R) \over (p^2)^{2}} 
+ O((Q^2)^3) ,
\end{align}
and  
\begin{align}
  D_T(-Q^2,\mu^2,g_R^2,\lambda_R) 
= - Q^2 \int_{0}^{\infty} dp^2 {\rho(p^2,\mu^2,g_R^2,\lambda_R) \over (p^2)^{2}} 
+ O((Q^2)^2) .
\end{align}
At first sight, this case is similar to the Gribov case. However, there is a crucial difference  and 
this case is unlikely to be realized as argued in the next subsection.

The above arguments hold as far as the IR divergence does not occur. 
In other words, the integrals for the spectral function are assumed to be finite.  This issue is discussed below.

\subsection{Moments of spectral function and mass gap}


\begin{figure}[htbp]
\begin{center}
\includegraphics[height=8cm]{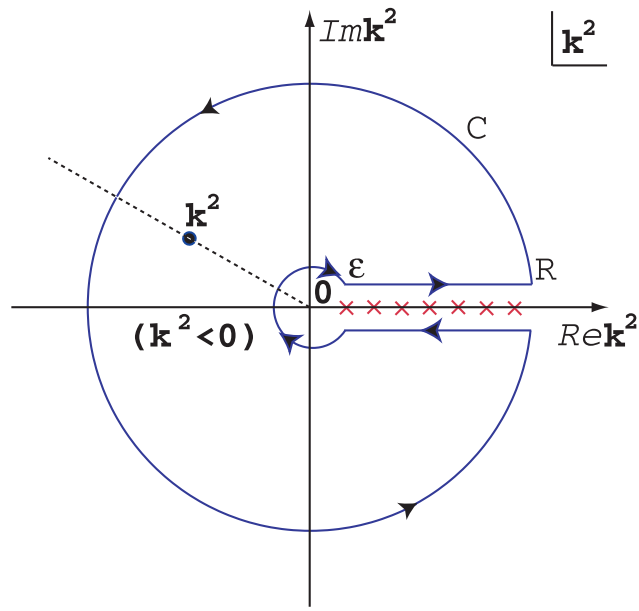}
\caption{The integration contour $C$ to which the Cauchy theorem is applied on the complex $k^2$ plane with singularities on the positive real axis.}
\label{fig:MomentContour}
\end{center}
\end{figure}


Now we consider the implication of the above result to the spectrum. 
To this end, we consider the moments of the spectral function:
\begin{align}
  \int_{\epsilon}^{R} ds s^{\ell} \rho(s) \quad (\ell= 0, \pm 1, \pm 2, \cdots) .
\end{align}
This integral is convergent for large $R$, if $\ell \le 1$ due to the asymptotic form (\ref{asymptoticrhoM}) of $\rho(s)$ for large $s=k^2$. Here any negative value of $\ell$ is allowed. 
In what follows, we pay attention to the convergence for the limit $\epsilon \rightarrow 0$.

To obtain the moment, we apply the Cauchy theorem to the function $f(z)$ which is analytic inside and on the contour $C$ in the complex $q^2$ plane shown in Fig.~\ref{fig:MomentContour} avoiding the positive real axis and the origin.  Then the function $z^{\ell}f(z)$  satisfies 
\begin{align}
  {1 \over 2\pi i} \oint_{C} dz \ z^{\ell} f(z) = 0 .
\label{Cauchy}
\end{align}
Here we separate this integral into three pieces; one is the contribution from the paths above and below the positive real axis $[\epsilon,R]$ which pick up the discontinuity  (the imaginary part) of the function $f(z)$  along the positive real axis; the others are the contribution to the integral over the circle of radius $R \gg 1$
and the contribution to the integral over the circle of radius $\epsilon \ll 1$. Hence (\ref{Cauchy}) is decomposed as
\begin{align}
  \int_{\epsilon}^{R} ds \ s^{\ell} {1 \over \pi}Im f(s+i0) 
  =& {1 \over 2\pi i} \oint_{|z|=\epsilon} dz \ z^{\ell} f(z)   - {1 \over 2\pi i} \oint_{|z|=R} dz \ z^{\ell} f(z)   
  \\
  =&  \int_{0}^{2\pi} {d\theta \over 2\pi}  \ \epsilon^{1+\ell} e^{i(1+\ell) \theta} f(\epsilon e^{i\theta})      -  \int_{0}^{2\pi} {d\theta \over 2\pi}  \ R^{1+\ell} e^{i(1+\ell) \theta} f(Re^{i\theta})    .
\label{Mom}
\end{align}
For $\ell=-1$, we apply (\ref{Mom}) to the gluon structure function 
$f=D_T$:
\begin{align}
  \int_{\epsilon}^{R} ds \ s^{-1} {1 \over \pi}Im D_T(s+i0) 
  =&  \int_{0}^{2\pi} {d\theta \over 2\pi}  \   D_T(\epsilon e^{i\theta})      -  \int_{0}^{2\pi} {d\theta \over 2\pi}  \   D_T(Re^{i\theta})   .
\label{n=-1}
\end{align}
In the region where the superconvergence relation holds, $D_T(z)$ vanishes in the limit $|z|=R \rightarrow \infty$. 
In the limit $R \rightarrow \infty$ and $\epsilon \rightarrow 0$, therefore, (\ref{n=-1}) reduces to  
\begin{align}
 \int_{0}^{\infty} ds \ s^{-1} \rho(s)
 \equiv  \int_{0}^{\infty} ds \ s^{-1} {1 \over \pi}Im D_T(s+i0) 
  =   \int_{0}^{2\pi} {d\theta \over 2\pi}  \   D_T(0) =  D_T(0)  ,  
\label{D(0)2}   
\end{align}
where  
 $\lim_{\epsilon \rightarrow 0}D_T(\epsilon e^{i\theta})$ has a common value in the limit of approaching the origin 
$|k^2|=\epsilon \rightarrow 0$ along the ray 
$k^2=|k^2|e^{i \theta} (0<\theta<2\pi)$ except for the positive real axis $\theta=0$. 
This result (\ref{D(0)2}) is consistent with the previous result (\ref{D(0)}). 

The finite $D_T(0)$ is consistent with the spectral function with a massive pole (at $p^2=M^2$):
\begin{align}
  \rho(s) = Z \delta(s-M^2) + \tilde{\rho}(s)      ,
  \label{specM}
\end{align}
where $\tilde{\rho}(s)$ is the contribution of the continuous spectrum from more than two particle states, 
namely, $\rho(s)$ has no support for $s<M^2$ (existence of the lower gap).
This is because the finiteness of the IR limit $D_T(0)$ denoted by the expression 
\begin{align}
  D_T(0) = \int_{0}^{\infty} ds \ s^{-1} \rho(s)
  = {Z \over M^2} + \int_{0}^{\infty} ds \ s^{-1} \tilde{\rho}(s)
\end{align}
is interpreted as indicating the existence of a massive pole ($M\not=0$), provided that $Z\not=0$.
\footnote{
It is possible to consider the absence of one-particle pole $Z=0$, i.e.,  $\rho(s) = \tilde{\rho}(s)$ in Minkowski region. 
However, we suppose that the asymptotic fields exist, even if the corresponding quanta can not be observed (i.e., they are confined). 
 Only in this formulation, we can adopt a sufficient condition for color confinement due to Kugo and Ojima.  Therefore, we restrict our consideration to the case in which this framework works.
} 
 In this sense, the finiteness of $D_T(0)$ excludes the massless gluon pole:
\begin{align}
  \rho(s) = Z \delta(s) + \tilde{\rho}(s)      .
\end{align}
The same argument holds for other moments of the spectral function appearing in the  power-series expansions (\ref{Fexpan}) and (\ref{Dexpan}).
Thus, we have shown that, for  gluon with  massive spectrum (\ref{specM}), then the power-series expansions (\ref{Fexpan}) and (\ref{Dexpan}) for the gluon propagator and the form factor could be well-defined for small Euclidean momenta $Q^2$.

Note that $0< D_T(0) < \infty$ corresponds to $\kappa=1/2$. 
This value for the IR exponent $\kappa$ seems to be supported by recent Monte Carlo simulations on a lattice \cite{Bonnetetal00}.
Quite recently, Bloch has proposed a truncation scheme respecting multiplicative renormalizability of the coupled Schwinger--Dyson equations for the gluon and ghost propagators by including all the diagrams \cite{Bloch03} to obtain the approximate solutions with the IR critical exponent $\kappa=1/2$.  

However, $\rho(s)$ is not necessarily positive for $s > 0$. Therefore, a situation of vanishing $D_T(0)$: 
$\int_{0}^{\infty} ds \ s^{-1} \rho(s)=D_T(0)=0$ happens to occur in principle.  
If this is the case, we must consider the next-to-leading term and then $\kappa=1$ will be realized (the apparent Gribov limit).  
However, we argue that $D_T(0)=0$ is unlikely to be realized based on the general properties of the spectral function.
 In fact, the structure function predicted by Gribov has the complex conjugate pair of poles
\begin{align}
  D_T(Q^2) = {Q^2 \over Q^4+M_G^4} 
= {1 \over 2} \left( {1 \over Q^2+iM_G^2} + {1 \over Q^2-iM_G^2} \right)  .
\end{align}
This is incompatible with the dispersion relation (\ref{EuclidD}) which requires the analyticity in the cut complex plane.

\subsection{IR behavior of the renormalized ghost propagator in Euclidean region}

For the Euclidean momenta $Q$, the ghost propagator has the spectral representation
\begin{align}
  \Delta_{FP}(Q^2, \mu^2, g_R^2, \lambda_R) 
 = \int_{0}^{\infty} dp^2 {\rho_{FP}(p^2, \mu^2, g_R^2, \lambda_R) \over p^2 +Q^2} 
=  {1 \over Q^2} G_{FP}(Q^2, \mu^2, g_R^2, \lambda_R) .
\end{align}
The propagator is not singular in Euclidean region $Q^2>0$. 
Hence we consider the IR limit $Q^2 \downarrow 0$. 
We see that the contribution from large $p^2$ to the integral  is convergent due to the asymptotic form of the spectral function derived in Appendix A.  
For ghost without the superconvergence relation,  
the IR limit of the ghost propagator is expected to be 
\begin{align}
  \Delta_{FP}(Q^2 \downarrow 0, \mu^2, g_R^2, \lambda_R) 
 = \int_{0}^{\infty} dp^2 {\rho_{FP}(p^2, \mu^2, g_R^2, \lambda_R) \over p^2} 
= \infty .
\label{ghost0}
\end{align}
If (\ref{ghost0}) was finite, the ghost form factor had  the similar form as the gluon case and the ghost form factor vanished in the IR limit:
\begin{align}
   G_{FP}(Q^2, \mu^2, g_R^2, \lambda_R) 
= Q^2 \int_{0}^{\infty} dp^2 {\rho_{FP}(p^2, \mu^2, g_R^2, \lambda_R) \over p^2} + O(Q^4) 
\rightarrow 0 \ \text{as} \ Q^2 \rightarrow 0 .
\end{align}
Eq.~(\ref{ghost0}) suggests that the ghost propagator can not have the power-series expansion in $Q^2$ around $Q^2=0$ in the Euclidean region.  
{}For the situation (\ref{ghost0}) to be realized, the ghost can not have the spectral function of massive type. 
This result does not contradict with the consideration based on the RG equation given below and the approximate solutions of the truncated coupled SD equations. 

If the ghost has the  massless spectrum of the usual type:
\begin{align}
 \rho_{FP}(p^2, \mu^2, g_R^2, \lambda_R) 
= Z_{FP} \delta(p^2) + \tilde{\rho}_{FP}(p^2) ,
\end{align}
with $Z_{FP}\not=0$,
then the ghost propagator reads
\begin{align}
  \Delta_{FP}(Q^2, \mu^2, g_R^2, \lambda_R) 
 = {Z_{FP} \over Q^2}
 + \int_{0}^{\infty} dp^2 {\tilde{\rho}_{FP}(p^2, \mu^2, g_R^2, \lambda_R) \over p^2 +Q^2} 
=  {1 \over Q^2} G_{FP}(Q^2, \mu^2, g_R^2, \lambda_R) .
\end{align}
This again contradicts with the consideration based on the RG equation below. 

Thus we reconsider the derivation of the dispersion relation given above. 
If the spectral function $\rho$ has singularities accumulating toward the origin $p^2=0$, we must replace an integration contour in Figure~\ref{fig:CutComplexPlane} by
a new contour in Figure~\ref{fig:MomentContour} to avoid the origin.
For a reference point $k^2$ inside a new contour $C$, the Cauchy integral formula yields  
\begin{align}
  f(k^2) =& {1 \over 2\pi i} \oint_{C} dz {f(z) \over z-k^2} 
  \\
  =&    {1 \over 2\pi i} \int_{0}^{R} dz {f(z+i\epsilon)-f(z-i\epsilon) \over z-k^2} 
\nonumber\\&
+ {1 \over 2\pi i} \oint_{|z|=R} dz {f(z) \over z-k^2} 
- {1 \over 2\pi i} \oint_{|z|=\epsilon} dz {f(z) \over z-k^2} .
\end{align}
 The similar argument leads to 
\begin{align}
  f(k^2) =&    {1 \over \pi} \int_{0}^{R} dz {Im f(z+i\epsilon) \over z-k^2} 
+   \int_{0}^{2\pi} {d\theta \over 2\pi} {Re^{i\theta} f(Re^{i\theta}) \over Re^{i\theta}-k^2} 
-   \int_{0}^{2\pi} {d\theta \over 2\pi} {\epsilon e^{i\theta} f(\epsilon e^{i\theta}) \over \epsilon e^{i\theta}-k^2} .
\end{align}
The second term vanishes for the function $f(z)$ which vanishes asymptotically as $|z| \rightarrow \infty$ and hence 
\begin{align}
  f(k^2) =    {1 \over \pi} \int_{0}^{R} dz {Im f(z+i\epsilon) \over z-k^2} 
+  \int_{0}^{2\pi} {d\theta \over 2\pi}  {\epsilon e^{i\theta} f(\epsilon e^{i\theta}) \over k^2- \epsilon e^{i\theta}} .
\end{align}
The effect of the last term is classified into three cases. 

i) In the limit $\epsilon \rightarrow 0$, the last term vanishes for the bounded function $f(z)$ which is regular in the limit $z \rightarrow 0$ (massive case).  Therefore, the above conclusion for the gluon with massive spectrum also holds for a new choice of the integration contour. 
This is an undesirable result. Therefore, we allow $f(z)$ to be singular in the limit $z \rightarrow 0$. 
For concreteness, we examine the asymptotic form 
$f(z) \sim C z^{\beta-1}$ for $|z| \ll 1$. 
 Nevertheless,  vanishing of the last term also follows for the function which becomes singular in such a way that $f(z) \sim C z^{\beta-1}$ with positive real $\beta>0$.  
This is also the case for a negative integer $\beta$, since  for sufficiently small $\epsilon$ 
\begin{align}
  \int_{0}^{2\pi} {d\theta \over 2\pi}  {\epsilon e^{i\theta} f(\epsilon e^{i\theta}) \over k^2- \epsilon e^{i\theta}} 
\cong {C\epsilon^\beta \over k^2} \int_{0}^{2\pi}  {d\theta \over 2\pi} e^{i\beta\theta}
= {C\epsilon^\beta \over k^2}  {e^{i\beta(2\pi)}-1 \over 2\pi i\beta} .
\end{align}

ii) For $\beta=0$, i.e., $f(z) \sim C z^{-1}$ (the tree massless form), however, we have the additional term:
\begin{align}
  f(k^2) =    {1 \over \pi} \int_{0}^{R} dz {Im f(z+i\epsilon) \over z-k^2} 
+ {C \over k^2}  .
\end{align}

iii) Finally, the additional term remains also for the singular function $f(z) \sim C z^{\beta-1}$ with {\it negative and non-integer }$\beta$ (IR enhanced case).
It is this case that corresponds to the ghost propagator in the Landau gauge, as argued below based on the RG equation.

\subsection{Renormalization condition}

This should be compared with the momentum subtraction scheme in perturbation theory where the renormalization condition fixes the gluon propagator to the tree-level one at a sufficiently large space-like renormalization point $k^2=\mu^2$, i.e.,
$D_T(\mu^2,\mu^2,g_R^2,\lambda_R)=1/\mu^2$. 
 This will lead to
\begin{align}
  \int_{0}^{\infty} dp^2  \rho(p^2,\mu^2 \rightarrow \infty,g_R^2 \rightarrow 0, \lambda_R \rightarrow  \lambda_*) = 1 .
\label{renorcond}
\end{align}
In the formal level, the spectral representation has the expansion for large $Q^2$
\begin{align}
  F(Q^2,\mu^2,g_R^2,\lambda_R) 
=& \int_{0}^{\infty} dp^2 {Q^2  \over Q^2+p^2} \rho(p^2,\mu^2,g_R^2,\lambda_R)
\nonumber\\
=& \sum_{n=0}^{\infty} {(-1)^{n} \over (Q^2)^n}
\int_{0}^{\infty} dp^2 (p^2)^n \rho(p^2,\mu^2,g_R^2,\lambda_R)   .
\end{align}
This power-series expansion in $1/Q^2$ is not well-defined, since finiteness of the integral as the coefficient is not guaranteed for $n >1$. 

The relation (\ref{renorcond}) seems to contradict with the superconvergence relation. However, the renormalization condition $F(\mu^2,\mu^2)=1$ is too strong, since it specifies in advance a part of the non-perturbative dynamics which should be determined by solving the theory.  
Therefore, we do not require the renormalization condition $F(\mu^2,\mu^2)=1$ even for large $\mu$ in this paper. 
Rather, we adopt the renormalization condition for any value of $\mu$
\begin{align}
  F(\mu^2,\mu^2)G^2(\mu^2,\mu^2)=1 ,
\end{align}
where $G$ is the ghost form factor. 
A loophole of the above argument is to take into account the FP ghost in the superconvergence relations.  

The superconvergence relation is consistent with
\begin{align}
 \lim_{Q^2 \rightarrow \infty} F(Q^2,\mu^2<\infty,g_R^2 >0, \lambda_R) 
= \int_{0}^{\infty} dp^2  \rho(p^2,\mu^2<\infty,g_R^2 >0, \lambda_R) 
= 0 .
\end{align}

Remarks:

1. The superconvergence relation is the spectral sum rule for the interacting theory.  
The superconvergence relation implies the violation of positivity of the spectral density of the transverse gluon propagator in Landau gauge.  
This can be interpreted as a manifestation of (gluon) confinement.  
However, it is not sufficient to conclude color confinement.  

2. In \cite{AS01}, it was argued that the superconvergence relation might therefore be interpreted as a reincarnation of Haag's theorem;  The free theory and the interacting theory are inequivalent no matter how small the coupling is.

\section{Infrared fixed point and color confinement}

The color confinement criterion of Kugo and Ojima is concerned with the IR behavior of the ghost form factor.  
Therefore, we focus on the IR behavior of the ghost form factor and its relationship with the gluon form factor.  

The gluon propagator for which the superconvergence holds can behave like a power of the momenta in the IR asymptotic region.  For the ghost, however, the superconvergence does not hold in general and hence we can not conclude the ghost propagator behaves like a power of the momenta in the IR region.  
The power-like solution of the gluon propagator is possible if and only if the RG function has the power series expansion in the coupling constant around an IR fixed point.  
In what follows, we argue that the power like behavior for both propagators is consistent with the RG functions which has the power series expansion in the coupling constant around an IR fixed point 
based on the RG equation.

\subsection{IR fixed point}

In what follows, we consider a phase of QCD which is connected to the asymptotic free regime, i.e., $\beta(0)=0$ and $\beta(\alpha)<0$ for sufficiently small $\alpha>0$.  
Let $\alpha_{\infty}$ be the infrared (IR) value of the running coupling constant $\alpha(\mu)$, i.e., $\alpha_{\infty}=\alpha(\mu=0)$.

\begin{figure}[htbp]
\begin{center}
\includegraphics[height=8cm]{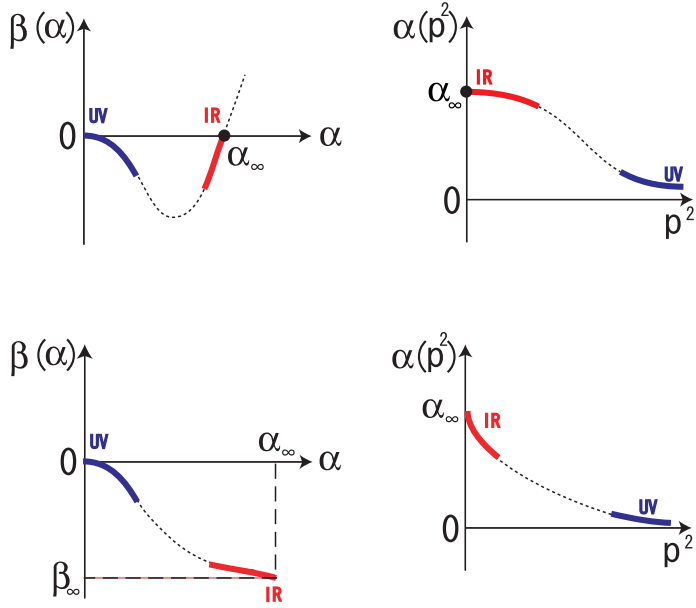}
\caption{Two types of the beta function and the corresponding running coupling constants.}
\label{fig:IRbetafunction}
\end{center}
\end{figure}

Suppose that the $\beta$-function is a continuous function of $g^2$.
If there is a region in which $\beta(\alpha)>0$, there must be at least one $\alpha_0$ such that $\beta(\alpha_0)=0$.
Therefore, we restrict our consideration to the first negative region  of $\beta(\alpha) \le 0$, i.e., $0 \le \alpha \le \alpha_0$ where 
 $\alpha_0$ is a zero closest to the origin.  See Figure~\ref{fig:IRbetafunction}.
If $\beta(\alpha)$ is negative in the whole region $\alpha>0$, we can interpret $\alpha_0=\infty$.


Now we proceed to examine which IR asymptotic forms of the propagator are possible as solutions of the RG equation. 
Suppose that the  RG functions $\beta(\alpha), \gamma(\alpha)$  are single-valued and continuous functions and that they can be expanded into the formal power series around $\alpha=\alpha_{\infty}$:
\begin{align}
 \beta(\alpha) =& \beta_{\infty} + (\alpha-\alpha_{\infty}) \beta_{1}^{IR} + {1 \over 2}(\alpha-\alpha_{\infty})^2 \beta_{2}^{IR} + \cdots ,\\
  \gamma(\alpha) =& \gamma_{\infty} + (\alpha-\alpha_{\infty}) \gamma_{1}^{IR} + {1 \over 2}(\alpha-\alpha_{\infty})^2 \gamma_{2}^{IR} + \cdots ,
\end{align}
where we have defined 
$\beta_\infty:=\beta(\alpha_\infty), \gamma_\infty=\gamma(\alpha_\infty)$. 

In our argument, the existence of the Landau pole 
(namely, the running coupling $\alpha$ blows up at a certain $\mu=\mu_0>0$ before reaching the IR limit $\mu=0$)
 is excluded from the beginning, since it is considered to be an artifact of the perturbation theory. 
We also excluded the case of $\beta_{\infty}=-\infty$, since the power series expansion around $\alpha=\alpha_{\infty}$ loses the meaning in this case. Even in this case, we consider that the anomalous dimension $\gamma_{\infty}$ remains finite $|\gamma_{\infty}|<\infty$ even in the IR limit. Otherwise, the field itself will lose its meaning in the IR limit. 

We need to estimate the ratio $\gamma(\alpha)/\beta(\alpha)$ to know the IR asymptotic form of the propagator where $\gamma(\alpha)$ depends on the field in question. 
We discuss two cases separately: (I) $\beta_{\infty}=0$, (II) $\beta_{\infty}\not=0 (<0)$.
In any case, the running of the coupling constant is governed by 
\begin{align}
  \int_{\alpha(\sigma)}^{\bar{\alpha}(p)} {d\alpha \over \beta(\alpha)} = \int_{\sigma}^{p} {d\mu \over \mu} .
\label{rel}
\end{align}

In the case of (I), the integration of (\ref{rel}):
\begin{align}
  \int_{\alpha(\sigma)}^{\bar{\alpha}(p)} {d\alpha \over (\alpha-\alpha_{\infty}) \beta_{1}^{IR}} \cong \int_{\sigma}^{p} {d\mu \over \mu} 
\end{align}
shows that the running coupling constant $\bar{\alpha}$ behaves for small $p$ as
\begin{align}
 {\bar{\alpha}(p)-\alpha_{\infty} \over \alpha(\sigma)-\alpha_{\infty}}
 \cong \left( {p \over \sigma} \right)^{\beta_{1}^{IR}} .
\label{scaling}
\end{align}
In this case, $\beta_{1}^{IR}>0$, since $\beta(0)=0$, $\beta(\alpha_\infty)=0$ and $\beta(\alpha)<0$ for $0<\alpha<\alpha_\infty$. 
Then, the ratio in the case (I) reads
\begin{align}
 {\gamma(\alpha) \over \beta(\alpha)} = {1 \over \alpha-\alpha_{\infty}} {1 \over \beta_{1}^{IR}} \left[ \gamma_{\infty} + (\alpha-\alpha_{\infty}) \left( \gamma_{1}^{IR}-{1 \over 2}\gamma_{\infty}{\beta_{2}^{IR} \over \beta_{1}^{IR}} \right) + O((\alpha-\alpha_{\infty})^2) \right] ,
\end{align}
which yields
\begin{align}
  \exp \left\{ 2\int_{\alpha}^{\bar{\alpha}} d\alpha {\gamma(\alpha) \over \beta(\alpha)}  \right\} 
  =  \left( {p^2 \over \sigma^2} \right)^{\gamma_{\infty}} 
  \exp \left\{ {(\alpha-\alpha_{\infty}) \over \beta_{1}^{IR}} \left( 2 \gamma_{1}^{IR}-\gamma_{\infty}{\beta_{2}^{IR} \over \beta_{1}^{IR}} \right)  [-1+(p/\sigma)^{\beta_{1}^{IR}}] \right\} ,
\end{align}
after rewriting $\bar{\alpha}(p)$ in terms of $p$ using (\ref{scaling}). Therefore, the form factor and the propagator of the $\Phi$-field  
behave respectively like 
\begin{align}
 F_{\Phi}(p^2) \cong \left( p^2/\sigma^2 \right)^{\gamma^{\Phi}_{\infty}} , \quad
 D_{\Phi}(p^2) := {F_{\Phi}(p^2) \over p^2} \cong \left( p^2/\sigma^2 \right)^{\gamma^{\Phi}_{\infty}-1} .
\label{powerlaw}
\end{align}

In the case (II), the IR behavior of 
the running coupling constant $\bar{\alpha}(p)$  is determined from 
\begin{align}
  \int_{\alpha(\sigma)}^{\bar{\alpha}(p)} {d\alpha \over \beta_{\infty}+(\alpha-\alpha_{\infty}) \beta_{1}^{IR}} = \int_{\sigma}^{p} {d\mu \over \mu} ,
\end{align}
which is integrated as
\begin{align}
 {\beta_{\infty}+(\bar{\alpha}(p)-\alpha_{\infty})\beta_{1}^{IR} \over \beta_{\infty}+(\alpha(\sigma)-\alpha_{\infty})\beta_{1}^{IR}}
 \cong \left( {p \over \sigma} \right)^{\beta_{1}^{IR}} .
\end{align}
Here the coefficient must be non-positive $\beta_{1}^{IR}\le 0$. Otherwise, we run into the contradiction. In fact, the RHS goes to zero as $p^2 \rightarrow 0$ for $\beta_{1}^{IR}>0$, while LHS is non-vanishing in this limit $\bar{\alpha}(p) \rightarrow \alpha_{\infty}$ for $\beta_{\infty}\not=0$.
On the other hand, the ratio in the case (II) given by
\begin{align}
 {\gamma(\alpha) \over \beta(\alpha)} = {1 \over \beta_{\infty}} \left[ \gamma_{\infty} + (\alpha-\alpha_{\infty}) \left( \gamma_{1}^{IR}-\beta_{1}^{IR}{\gamma_{\infty} \over \beta_{\infty}} \right) + O((\alpha-\alpha_{\infty})^2) \right] 
\end{align}
yields
\begin{align}
  \exp \left\{ 2\int_{\alpha}^{\bar{\alpha}} d\alpha {\gamma(\alpha) \over \beta(\alpha)}  \right\} 
  =   
  \exp \left\{ 
  (\bar{\alpha}-\alpha){\gamma_{\infty} \over \beta_{\infty}} + {1 \over 2}[(\bar{\alpha}-\alpha_{\infty})^2-(\alpha-\alpha_{\infty})^2]
  {1 \over \beta_{\infty}}
 \left( \gamma_{1}^{IR}-\beta_{1}^{IR}{\gamma_{\infty} \over \beta_{\infty}} \right) 
  \right\} .
\end{align}
This expression converges to a constant in the IR limit $p^2 \rightarrow 0$.  Therefore the leading IR power behavior of the propagator is still given by the tree form: $D_\Phi(p^2) \cong \text{const.}(p^2)^{-1}$ in the case (II).

Thus we conclude that the second case (II) is excluded for the gluon by the superconvergence relation, suggesting the existence of the IR stable fixed point $\beta_{\infty}:=\beta(\alpha_{\infty})=0$. 
Since the $\beta$-function is common to all the fields, the other fields should obey the power law (\ref{powerlaw}) but with different anomalous dimensions $\gamma_{\infty}^{\Phi}$.%
\footnote{
Our argument should be compared with that of \cite{WA01} 
(See also Appendix A in \cite{LS02} for some minor corrections) where it is pointed out that the power-series in momentum (without logarithmic corrections \cite{Kondo02a})
$
 G(Q^2, \mu^2) = \sum_{n} d_n(g) (Q^2/\mu^2)^{\delta_n} 
$
 can be a solution of the RG equation for the ghost formfactor in the Landau gauge $\lambda=0$ 
if and only if the exponents $\delta_n$ are $g$ independent and 
$
 d_n(g) \cong \exp \left\{ \int^{g} dl {2[\delta_n+\gamma(l)] \over \beta(l)} \right\}
$.
Here the RG functions $\beta(l)$ and $\gamma(l)$ are unspecified in sharp contrast with our consideration.
The exponent $\delta$ was determined based on the approximate solution of the SD equation for gluon and ghost propagators.  
}
For the gluon, we use 
\begin{align}
  \gamma_{\infty}^{A}=2\kappa=1 
\end{align}
  from the above argument based on the superconvergence.  
The remaining ghost anomalous dimension $\gamma_{\infty}^{C}$ is determined as follows.

The non-renormalization \cite{Taylor71} of gluon--ghost--antighost vertex in the Landau gauge, i.e., 
\begin{align}
 \tilde{Z}_1 \equiv Z_3^{1/2} \tilde{Z}_3 Z_g = 1
\end{align}
yields, by taking the logarithm and operating the differential operator $\mu {\partial \over \partial \mu}$, an identity among the RG functions 
\begin{align}
  \gamma^{A}(\alpha)+2\gamma^{C}(\alpha)
= {\beta(g) \over g}
 ={\beta(\alpha) \over 2\alpha} \le 0 .
\label{RGrelation}
\end{align}
Then we have a relationship in the Landau gauge in the case (I) at 
$\alpha=\alpha_{\infty}$
\begin{align}
 \gamma_{\infty}^{A}+2\gamma_{\infty}^{C}=0 .
\label{RGrelation2}
\end{align}
Therefore, the sign of $\gamma_{\infty}^{C}$ is opposite to $\gamma_{\infty}^{A}$.
Substituting the anomalous dimension of the gluon $\gamma_{\infty}^{A}=1$ into (\ref{RGrelation2}), we obtain 
\begin{align}
 \gamma_{\infty}^{A}=1, \quad \gamma_{\infty}^{C}=-1/2 
\end{align}
 in the Landau gauge  in the case (I).
Thus, the gluon propagator is suppressed while the ghost propagator is enhanced in the IR region.

However, this result seems to be incompatible with the superconvergence of the ghost in the Landau gauge, which may yield  
$\gamma_{\infty}^{C}=1$ for the massive ghost. 
Note that the Landau gauge $\lambda=0$ is the fixed point, but it is the IR stable fixed point and not the UV stable fixed point.  For superconvergence, we need the behavior of the UV limit where the perturbation theory is reliable due to asymptotic freedom.  
The superconvergence for the ghost does not hold for $\lambda \not= 0$, no matter how small $\lambda$ is.  The result depends on the initial value of $\lambda$. 
Therefore, the treatment of the superconvergence of the ghost propagator is more subtle than the gluon propagator.
The superconvergence of ghost in the Landau gauge is very special compared with that of gluon which has superconvergence irrespective of the choice of $\lambda$.
As we have shown in the previous section, this difficulty can be avoided if the ghost does not have the massive spectrum and the power-series expansion loses the meaning.  
In fact, the above consideration based on the RG equation suggests that the ghost propagator is enhanced in the IR and of the long-range type, while the gluon propagator is suppressed in the IR and of the short-range type.%
\footnote{
As a theoretical possibility, we can not exclude the converse case based on this argument alone, namely,
the gluon propagator is enhanced in the IR and of the long-range type, while the ghost propagator is suppressed in the IR and of the short-range type.  However, such a solution with $\kappa<0$ can not be a consistent solution of the truncated coupled SD equations for the gluon and ghost propagators. 
}

\subsection{Color confinement}

A sufficient condition for color confinement was proposed by Kugo and Ojima (KO) \cite{KO78,KO79}, see also \cite{Kugo89,NO90}.  
The criterion in the original form is a condition 
$u(0)=-1$
for the two-point function of the composite operators: 
\begin{align}
 \delta^{AB}  
\left( g_{\mu\nu} - {p_\mu p_\nu \over p^2} \right) u(p) 
 := \int d^4x e^{ip(x-y)} \langle 0| {\rm T} (D_\mu C)^A(x) g(A_\nu \times \bar{C})^B(y)|0 \rangle  .
\label{KOparameter1}
\end{align}
In particular, it was shown later that the criterion is considerably simplified in the Landau gauge $\lambda=0$ \cite{Kugo95}. 
It was reduced to a condition for the ghost form factor in the IR limit:
\begin{equation}
 \lim_{Q^2 \rightarrow 0} [Q^2 \Delta_{FP}(Q^2)]^{-1} 
\equiv  \lim_{Q^2 \rightarrow 0} [G(Q^2)]^{-1}
\equiv 1+u(0) = 0 .
\label{KO}
\end{equation}
Our argument given above leads to the ghost form factor enhanced in the IR:
\begin{equation}
  G(Q^2)  \cong \left( Q^2/\sigma^2 \right)^{\gamma^{C}_{\infty}} 
\cong \left( Q^2/\sigma^2 \right)^{-1/2} .
\end{equation}
Therefore, the solution found in this article satisfies the KO color confinement criterion  (\ref{KO}). 
In this sense, color confinement is achieved in the Landau gauge Yang-Mills theory.%

As demonstrated in this article, the superconvergence alone is not sufficient to draw the conclusion for color confinement.  
In fact, the KO criterion needs the property of the IR limit $Q^2 \rightarrow 0$ of the two-point function for ghost and antighost. 
In the gauges other than the Landau, the detailed information for gluon is also needed to verify the KO criterion.  
According to recent investigations of the coupled SD equations for gluon and ghost, contribution from ghost plays the dominant role in the IR region.

This result should be compared with the Nishijima criterion \cite{Nishijima86,Nishijima87,Nishijima94,CN01} for color confinement 
\begin{equation}
  Z_{3}^{-1} = 0 .
\end{equation}
This criterion is also satisfied in this case, since the combination of 
the renormalized spectral representation and the renormalized superconvergence relation implies \begin{align}
  Z_{3}^{-1} = \int_{0}^{\infty}  dp^2 \rho(p^2, \mu^2, g_R^2, \lambda_R) = {\lambda \over \lambda_{*}} ,
\end{align}
as already argued by himself.  

Our argument just given is not complete due to an assumption on the RG functions, as the following example shows.

\subsection{IR attractor}


\begin{figure}[htbp]
\begin{center}
\includegraphics[height=5cm]{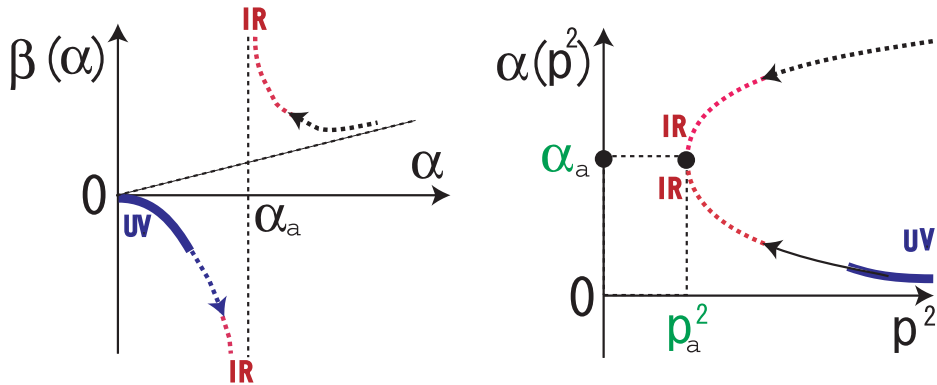}
\caption{The beta function with an IR attractor and the corresponding running coupling constant.}
\label{fig:IRattractor.eps}
\end{center}
\end{figure}


Moreover, there is an exceptional case to the above argument. 
It is the case of IR attractor where $\beta$ function has a pole (not a zero) at a certain point $\alpha_a$, i.e., $\beta(\alpha_a-0)=-\infty$, see Figure~\ref{fig:IRattractor.eps}.
In this case, the $\beta$-function blows up (down) at $\alpha_a$ before reaching the IR limit $\alpha_{\infty}$ and hence the $\beta$-function cannot be expanded around $\alpha=\alpha_{\infty}$ (Note that $\alpha_{\infty}>\alpha_a$). 
Some authors \cite{CEMS00} argued that the IR attractor will be realized in QCD with  small number of flavors, just as the supersymmetric (SUSY) gluodynamics has the IR attractor \cite{NSVZ83,KS95}.
The exact NSVZ \cite{NSVZ83} $\beta$ function for SUSY gluodynamics is given by
\begin{align}
 \beta(\alpha) = - {\alpha^2 \over 2\pi} {3C_2(G) \over 1-C_2(G) \alpha/2\pi} ,
\end{align}
where $C_2(G)$ is the quadratic Casimir operator in the adjoint representation, $C_2(G)=N_c$ for $SU(N_c)$. 
Of course, it has the power series expansion around $\alpha=0$:
\begin{align}
 \beta(\alpha) = - {\alpha^2 \over 2\pi}  3C_2(G)  \sum_{n=0}^{\infty} \left( C_2(G) {\alpha \over 2\pi} \right)^{n} .
\end{align}
In the case of IR attractor, the coupling constant $\alpha(\mu)$ becomes complex-valued below the scale $\mu_a$ at which the pole exists $\alpha_a=\alpha(\mu_a)$. Therefore, we need independent study in the case of IR attractor.

\section{MA gauge}

We have obtained all the possible superconvergence relations based on the asymptotic behavior of the propagator. See Appendix~A for calculations. 
Summarizing the result, the superconvergence relation holds for the diagonal gluon, off-diagonal FP ghost, off-diagonal gluon, while the superconvergence does not occur for the diagonal FP ghost.

In what follows, we choose the gauge parameter $\beta=0$ for diagonal components and omit to write $\beta=0$ hereafter. The the following unsubtracted dispersion relations for the propagators and form factors and the superconvergence relations hold irrespective of the gauge parameter $\alpha$ for off-diagonal components. 
The diagonal gluons have
\begin{align}
  D_{diag}(k^2, \mu^2, g^2, \alpha) 
 =& \int_{0}^{\infty} dp^2 {\rho_{diag}(p^2, \mu^2, g^2, \alpha) \over p^2 -k^2} ,
\\
  F_{diag}(k^2, \mu^2, g^2, \alpha) =&  - \int_{-0}^{\infty} dp^2  {p^2 \rho_{diag}(p^2, \mu^2, g^2, \alpha) \over p^2 -k^2} ,
\\
 0 = & \int_{0}^{\infty} dk^2 \rho_{diag}(k^2, \mu^2, g^2, \alpha) .  
\end{align}
The off-diagonal gluons have  
\begin{align}
  D_{A}(k^2, \mu^2, g^2, \alpha) 
 =& \int_{0}^{\infty} dp^2 {\rho_{A}(p^2, \mu^2, g^2, \alpha) \over p^2 -k^2} ,
\\
  F_{T}(k^2, \mu^2, g^2, \alpha) =& - \int_{-0}^{\infty} dp^2  {p^2 \rho_{A}(p^2, \mu^2, g^2, \alpha) \over p^2 -k^2} ,
\\
  0 = &  \int_{0}^{\infty} dk^2 \rho_{A}(k^2, \mu^2, g^2, \alpha) .
\end{align}
The off-diagonal FP ghosts have
\begin{align}
  \Delta_{FP}(k^2, \mu^2, g^2, \alpha) 
 =& \int_{0}^{\infty} dp^2 {\rho_{FP}(p^2, \mu^2, g^2, \alpha) \over p^2 -k^2} ,
\\
  G_{FP}(k^2, \mu^2, g^2, \alpha) =&   - \int_{-0}^{\infty} dp^2  {p^2 \rho_{FP}(p^2, \mu^2, g^2, \alpha) \over p^2 -k^2} ,
\\
  0 = & \int_{0}^{\infty} dk^2 \rho_{FP}(k^2, \mu^2, g^2, \alpha) .
\end{align}

However, the diagonal FP ghost propagator has an unsubtracted dispersion relation  
\begin{align}
  \Delta_{diag}(k^2, \mu^2, g^2, \alpha) 
 = \int_{0}^{\infty} dp^2 {\rho_{c}(p^2, \mu^2, g^2, \alpha) \over p^2 -k^2} 
= -{1 \over k^2} G_{diag}(k^2, \mu^2, g^2, \alpha) 
\end{align}
without superconvergence.

Therefore, the diagonal gluon, off-diagonal ghost and off-diagonal gluon become massive.  
 From the viewpoint of Abelian dominance, the mass $m_a$ of the diagonal gluon must be smaller than masses $m_A, m_C$ of the off-diagonal gluon and off-diagonal ghost.  
The off-diagonal propagator must be short-ranged compared to the diagonal one.  
Indeed, a numerical simulation on a lattice indicates this is the case \cite{AS99}. Moreover, the latest simulation has confirmed this result and provided more detail, i.e., $m_A \cong 2 m_a$ \cite{BCGMP03} where the gauge parameters adopted correspond to  $\alpha_R=0$ and $\beta_R=0$.
It should be remarked that the ghost propagators have not yet been measured in these lattice studies \cite{AS99,BCGMP03}. 
The above results show that the off-diagonal ghost propagator 
We can not say anything about the diagonal ghost.  However, only the diagonal ghost propagator can have the long-range correlation.  
This is necessary for the confinement to hold, since the only the long-range correlation can convey the information to the long distance region where confinement is relevant.  

The massiveness of the diagonal gluon is unexpected at first sight.  
This is because the residual U(1) invariance seemed to impose the masslessness of the diagonal gluon and the long-range mode is needed to explain quark confinement, e.g., the linear potential.
However, we must recall that the ghosts play the role of the longitudinal and scalar gluons in the Landau gauge where the longitudinal and scalar gluon disappear. 
Therefore, the linear potential will be provided by the contribution from the diagonal ghost.  
In addition, the massive diagonal gluon after gauge fixing does not contradict the general principle of quantum field theory as discussed below form the viewpoint of the residual local gauge symmetry. 
This scenario seems to be not inconsistent with the confinement picture obtained in the case of Lorentz gauge fixing.  All the gluons are expected to become massive, while all the ghosts are expected to be long range, provided that the color symmetry is unbroken.  
In the MA gauge, the color symmetry is explicitly broken by the gauge fixing procedure.  Therefore, the diagonal and off-diagonal components can behave differently.

\section{Conclusion and discussion}

In this article we have examined how the propagators and the form factors for gluon and ghost behave as complex-valued functions on the complex $p^2$ plane from the viewpoint of general principles (multiplicative renormalizability, analyticity, Poincar\'e covariance, spectral condition, completeness) of quantized gauge field theories formulated in the manifestly Lorentz covariant fashion. 
In particular, we have emphasized that it is the analyticity of the propagator and the corresponding form factor on the complex $p^2$ plane that enables us to connect the UV asymptotic solution into the IR asymptotic solution and vice versa. 

We have obtained all the possible superconvergence relations for   gluons and ghosts in the generalized Lorentz gauge and the modified MA gauge, by making use of their UV asymptotic behaviors which are calculable reliably by perturbation theory due to asymptotic freedom. 
The superconvergence relations are obtained by combining two sum rules for a propagator and an associated form factor following  from the analyticity on the cut complex $p^2$ plane. 
Thereby, we can predict the IR asymptotic behavior of the Euclidean field propagator for which the superconvergence relation holds. 
By equating the dispersion relation following from analyticity with the spectral representation following  from the spectral condition and Poincar\'e covariance, it has been shown that the propagator with superconvergence may have massive pole and hence become short-range.  
Moreover, we have discussed the implications of superconvergence to color confinement in both gauges.  In other words, we have presented a scenario of deriving color confinement.  

In the generalized Lorentz gauge, we have confirmed that the gluon has superconvergence for arbitrary gauge parameter and hence the gluon propagator may become short-range.  
However, {\it the superconvergence for the ghost should not hold for arbitrary gauge}. %
\footnote{
In the manifestly Lorentz covariant formulation, even the spectral function $\rho(p^2)$ of the transverse gluon is no longer positive.  In fact, it is negative for large $p^2$ due to the indefinite metric of state space. In view of this, it is difficult to give an intuitive interpretation to the superconvergence relation from the physical point of view.  
In the generalized Lorentz gauge, the longitudinal part including the gauge parameter is also renormalized, in sharp contrast with the usual Lorentz gauge.  Especially, in the Landau gauge, the gluon propagator does not have the longitudinal part.  The unphysical degrees of freedom such as the longitudinal and the scalar parts are related to the ghost and antighost that are also unphysical, but exist even in the Landau gauge. 
 Therefore, we have no reason to distinguish the ghost from the other unphysical components of the gluon.  
}
This is  consistent with the approximate solutions of the coupled SD equations for gluons and ghosts in the Landau gauge \cite{SHA97,AB98,AB98b,Bloch01,Bloch02,FAR02,LS02,Kondo02a,Kondo02b,Bloch03}. If the superconvergence held for the ghost, the ghost propagator would become short-range.  
However, it is reasonable that the IR enhanced ghost propagator be long-range, since only the IR enhanced ghost could be a candidate for a carrier of confinement.
 We have concluded that {\it the gluon propagator exhibits the power behavior characterized by the  IR critical exponent  $\kappa=1/2$,
 \begin{equation}
 D_T(Q^2) \cong \text{const.}   + O(Q^2), 
\label{IRsol1}
\end{equation}
 while the ghost propagator has  a negative and non-integer exponent}.

Supposing the RG functions have the power series in  the coupling constant $g^2$ around the IR value $g^2(\mu=0)$, we have specified the IR asymptotic behavior for the gluon and ghost propagators so as to be consistent with the above consequence of superconvergence relations.
Moreover, we required that the solution is compatible with the existence of the IR fixed point.
Then the IR critical exponent of the ghost is determined. 
Thus we conclude that the  ghost propagator in the Landau gauge behaves like %
\footnote{
If the constant is zero in the gluon propagator (\ref{IRsol1}), i.e., 
$D_T(Q^2) \cong O(Q^2) \downarrow 0$, 
the   the ghost propagator behaves like
$ \Delta_{FP}(Q^2) \cong (Q^2)^{-2} \uparrow \infty$ instead of (\ref{IRsol2}).
This is nothing but the Gribov limit, $\kappa=1$ \cite{Gribov78}.
However, the explicit form obtained by Gribov contradicts with the analyticity requirement, since the Gribov form has a complex conjugate pair of poles. 
}
\begin{equation}
 \Delta_{FP}(Q^2) \cong (Q^2)^{-3/2} .
\label{IRsol2}
\end{equation}
This result is also consistent with the SD equation approach, if and only if the IR critical exponent $\kappa$ takes the value $\kappa=1/2$.
The IR asymptotic behavior of the ghost propagator (\ref{IRsol2}) fulfills a sufficient condition of color confinement in the Landau gauge presented in \cite{Kugo95} which is originally due to Kugo and Ojima \cite{KO79}, see also \cite{Kugo89,NO90}.  
The negative half integer exponent for the ghost implies the existence of the branch cut beginning at the origin and extending on the positive real axis in the complex $k^2$ plane.  
This is indeed the situation which is necessary to adopt a new integration contour in discussing anew the superconvergence relation for the ghost.

\begin{figure}[htbp]
\begin{center}
\includegraphics[height=4cm]{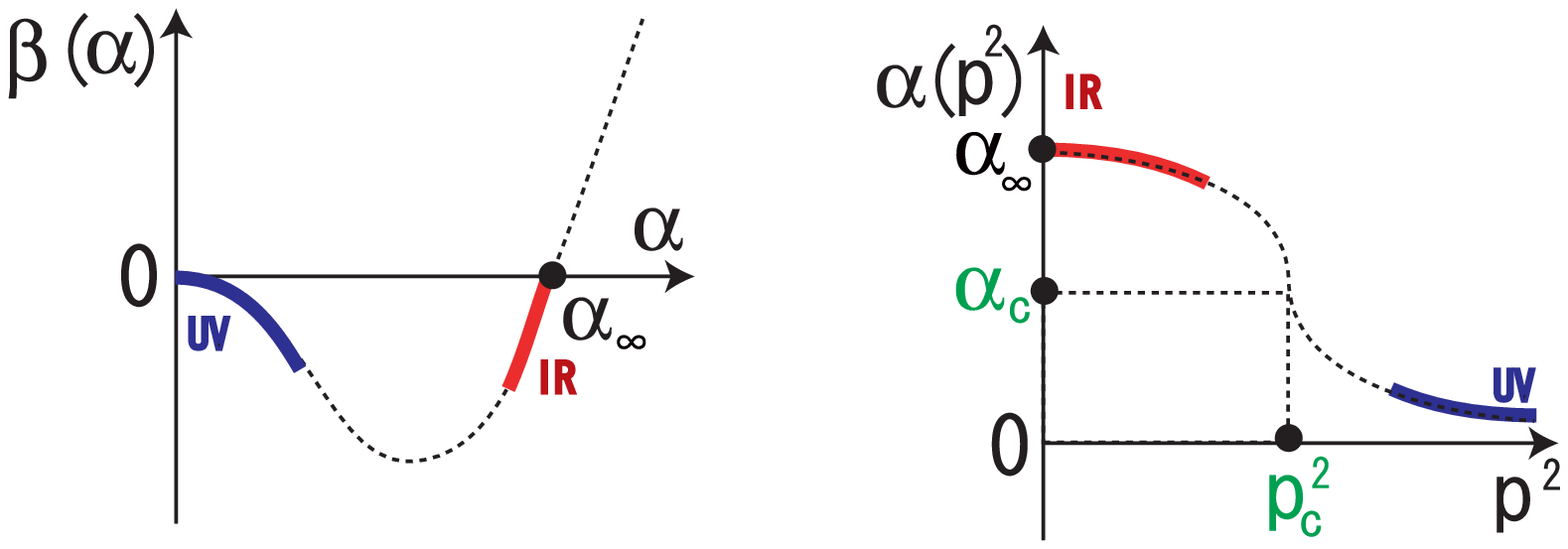}
\includegraphics[height=4cm]{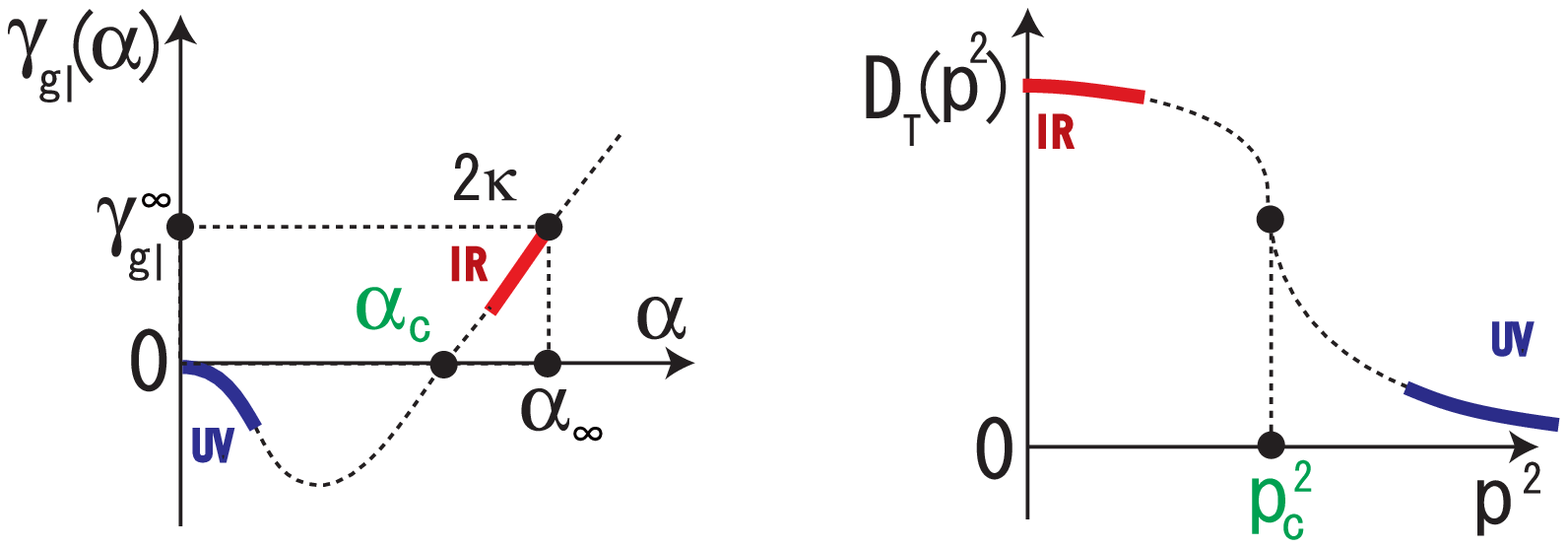}
\includegraphics[height=4cm]{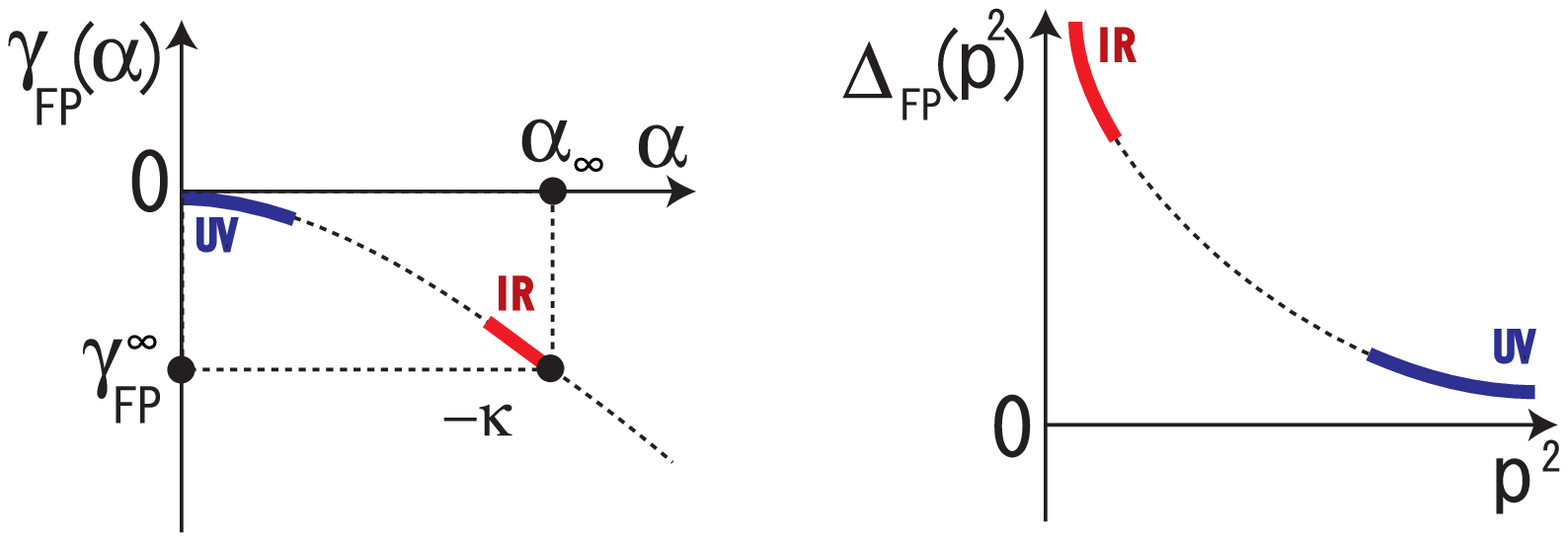}
\caption{
The schematic behavior of RG functions, running coupling constant and propagators for gluons and ghosts for Yang--Mills theory in the Lorentz--Landau gauge. 
(a) Beta function vs. Running coupling,
(b) Anomalous dim. of gluon vs. Gluon propagator,
(c) Anomalous dim. of ghost vs. FP Ghost propagator.
}
\label{fig:RGfunct-pfunct2}
\end{center}
\end{figure}

The IR limit of the running coupling constant $g^2(0)$ can be estimated by solving the ghost SD equation by substituting the IR asymptotic solutions (\ref{IRsol1}) and (\ref{IRsol2}), as already worked out in the SD equation approach, see e.g. Appendix A of \cite{Bloch02}.   
In Figure~\ref{fig:RGfunct-pfunct2}, we have summarized the relationship between the RG functions ($\beta$ function and anomalous dimension) and the running coupling constant and propagators.
An exception to be above argument is the solution corresponding to the IR attractor, which has been suggested to exist for QCD with quark flavors below a certain critical number based on the higher order analysis of perturbative series \cite{CEMS00}.  This case is to be investigated separately. 


 Our results  do not contradict with the existence of the gluon condensate 
 $\langle \mathscr{A}_\mu^2 \rangle$ of mass dimension two.
The reason is as follows. The superconvergence relation has been derived using only  the leading term of the ultraviolet asymptotic form of the propagator, while the relevant condensate appears as the sub-leading term of the ultraviolet asymptotic propagator, as the OPE calculation of the gluon propagator shows.  Therefore, the existence of such a condensate of mass dimension two does not influence the existence or non-existence of the superconvergence relation.  [In a very special situation (if any) in which all the perturbative terms vanish, such a condensate will be directly related to the infrared behavior of the propagator through the analyticity.] 
This is also the case for the gluon--ghost condensate \cite{Kondo01,KMSI02,Kondo03b} in the Maximal Abelian gauge.

It should be remarked that our results hold for arbitrary choice of gauge fixing. Therefore, the IR enhancement of the ghost propagator should hold for both the Landau and non-Landau gauges in the Lorentz type gauge fixing.  This is also the case for arbitrary gauge fixing parameters in the Curci-Ferrari gauge. 
Moreover, the IR enhancement of the ghost propagator in the Curci-Ferrari gauge is compatible with the existence (or finiteness) of the on-shell BRST-invariant gluon-ghost condensate 
$
  \langle {1 \over 2} \mathscr{A}_\mu \cdot \mathscr{A}_\mu  +   \lambda i \bar{\mathscr{C}} \cdot  \mathscr{C} \rangle 
$
which is written in terms of the gluon and ghost propagators and their form factors as (see (\ref{gluonpropa}) for the definition)
\begin{align}
  \Big\langle {1 \over 2} \mathscr{A}_\mu \cdot \mathscr{A}_\mu  +   \lambda i \bar{\mathscr{C}} \cdot  \mathscr{C} \Big\rangle 
=& (N^2-1) \int {d^4Q \over (2\pi)^4} \left[  {1 \over 2}D_{\mu\mu}(Q^2)   + \lambda  \Delta_{FP}(Q^2) \right] 
\nonumber\\
=& (N^2-1) \int_{\epsilon^2}^{\Lambda^2} {dQ^2 \over 16\pi^2}    \left[  {3 \over 2}F(Q^2) + {\lambda \over 2} F_L(Q^2) + \lambda  G(Q^2) \right] .
\end{align}
{}For the form factors having the IR power-law behaviors  
\begin{equation}
  F(Q^2) \cong A \cdot (Q^2)^{2\kappa} , \quad 
  F_L(Q^2) \cong A' \cdot (Q^2)^{-\kappa'} , \quad 
 G(Q^2) \cong B \cdot (Q^2)^{-\kappa}  ,
\end{equation}  
the integration of the last term is convergent for arbitrary $\lambda$ (after renormalization of the UV divergence), if $\kappa < 1$ (excluding the Gribov limit $\kappa=1$ again), since it behaves as $(\epsilon^2)^{1-\kappa}$ for the lower bound $\epsilon$. 
Therefore, the dominant contribution to the gluon--ghost condensate comes from the UV asymptotic region, even if the ghost propagator is enhanced in the IR limit.%
\footnote{For the condensate 
$
  \langle {1 \over 2} \mathscr{A}_\mu \cdot \mathscr{A}_\mu  \rangle 
$
in the usual Lorentz gauge, the lattice simulation exhibits the dominance of the UV contribution for the integral of $F(Q^2)$  in the Landau gauge $\lambda=0$, see \cite{Boucaudetal02b}.}  
The similar argument can be given also in the MA gauge. 

It is also expected that there exists some relationship between the vacuum condensate and the non-trivial topological object. 
However, the superconvergence relation utilizes only the leading term of the RG improved perturbation theory to characterize the ultraviolet asymptotic behavior which is continued analytically to the infrared region.  The non-trivial topological object, e.g., instanton   clearly appears in the non-perturbative treatment. 
Therefore, it is beyond the present framework of this article to extract or separate the effect of the topological object.  
This issue is to be tackled again in the near future.

In the MA gauge, we have shown that the off-diagonal gluon and the  off-diagonal ghost have superconvergence and hence all the off-diagonal propagators converge to $Q^2$-independent constants in the IR limit.  This implies that the off-diagonal modes become massive in the MA gauge as  is expected from the Abelian dominance \cite{tHooft81,EI82,SY90}.
This results do not contradict with the existence of the vacuum condensates of mass dimension 2 in the MA gauge \cite{Kondo01,Kondo03b}. 
We have found that {\it the diagonal gluon propagator may be of the massive type}, since the superconvergence holds. 
This is a novel feature discovered for the first time in our analytical study, supporting the numerical study \cite{BCGMP03}.
This implies that {\it the diagonal gluon becomes massive without breaking the residual local U(1) invariance}.
We wish to emphasize the importance of the diagonal ghost propagator in pure Yang--Mills theory with the MA gauge fixing. 
We conjecture that only the diagonal ghost propagator can be short-range to mediate the long-range confining force in the manifestly covariant formulation, since only the diagonal ghost does not obey superconvergence. 
 These results can be compared with numerical results  \cite{AS99,BCGMP03} where the ghost propagators are not measured. 
There is another possibility in the MA gauge, as will be discussed in a forthcoming article. 

In the MA gauge, therefore, the SD equation approach in addition to numerical simulations is indispensable to obtain the precise behavior of propagators, especially, the diagonal gluon and diagonal ghost propagators as in the Lorentz--Landau gauge. 
In this study, it is very important to reproduce correctly the one-loop perturbation result as a self-consistent solution in the asymptotic UV region.  Consequently, the analyticity allows the UV asymptotic solution to be analytically continued into the  correct IR asymptotic solution.  This requirement will help to constrain the approximation (for the vertices) to be adopted for writing down the solvable SD equation in MA gauge.  At the same time, the truncation of the SD equation must be performed in consistent with the multiplicative renormalizability.  
These points could be good criteria of judging the efficiency of the adopted approximations.  
The SD equations of gluodynamics in MA gauge will be discussed in a separate article, see \cite{KMS03} for a preliminary result.

 From our viewpoint, recent studies of the coupled SD equations for gluon and ghost propagators in the Landau gauge performed by Bloch \cite{Bloch01,Bloch02,Bloch03} are very interesting in the following respects.
He has proposed the truncation scheme respecting the multiplicative renormalizability and reproducing the leading order resumed perturbative results (beyond one-loop result) in the UV region \cite{Bloch01}.
Then he has actually solved the coupled SD equations for gluon and ghost propagators by including for the first time  all the diagrams (including two-loop diagrams in the gluon equation) and has found the numerical solution over the whole range of momenta characterized by the power behavior in the IR region \cite{Bloch03}.  
Among other things, his solution includes the case of $\kappa=0.5$, although all the approximate solutions obtained so far has $\kappa$  slightly larger than 0.5.
  His solution seems to be an approximate solution which is consistent with the general principles of gauge field theories.  

Here we comment on whether or not our results are affected by the Gribov problem. 
Gribov \cite{Gribov78} studied how the requirement of preserving  positivity of the ghost propagator influence the original Yang--Mills theory. 
This requirement forces one to restrict the region of functional integration over the gauge field to a subspace connected to the origin (called the first Gribov region) inside the Gribov horizon.  
Our results still hold even in the presence of Gribov problem,  
as far as  the analyticity is kept even after  restricting the functional integration.
Otherwise, it becomes impossible to perform the analytic continuation of the Euclidean propagator back to the Minkowski region.

In order to know the analytic structure of the propagators more explicitly, it is desirable to solve the SD equations on the complex $p^2$ plane.  Such investigations have been performed only for the quark SD equation \cite{SC90,MH92}.  
Solving the coupled SD equations for gluon and ghost on complex $p^2$ plane will  be a challenge in future.  
See  \cite{Aisoetal97} for numerical simulations. 
In a subsequent article we will discuss the effect of including the fermion flavors and clarify the interplay between color confinement/deconfinement and chiral symmetry breaking/restoration. 


\section*{Acknowledgments}
The author would like to thank Daniel Zwanziger for correspondences,  Maxim Chernodub for discussions and correspondences, and Jacques C.R. Bloch for private communications and sending the draft of the article \cite{Bloch03} prior to publication. 
He also would like to thank Izumi Ojima for helpful conversations and encouragements, Reinhardt Alkofer, Taichiro Kugo, Hideo Nakajima, Sadataka Furui, Hisao Suzuki,  and  Anthony G. Williams for helpful conversations on the related subjects. 
He is grateful to Takahito Imai for drawing some of figures. 
This work is supported by Sumitomo Foundations, Grant-in-Aid for Scientific Research on Priority Areas (B)13135203 from
The Ministry of Education, Culture, Sports, Science and Technology (MEXT), and 
Grant-in-Aid for Scientific Research (C)14540243 from Japan Society for the Promotion of Science (JSPS).

\appendix
\section{Calculation of UV asymptotic behavior of the propagators}

\subsection{Generalized Lorentz gauge (Curci--Ferrari gauge)}

The running of the gauge coupling constant is governed by the differential equation:
\begin{align}
   \beta(g_{\rm R}^2) := \mu {\partial g_{\rm R}^2 \over \partial \mu} 
 =-  {11 \over 3}C_2(G) {2g_{\rm R}^4 \over (4\pi)^2}  ,
\label{betaf2}
\end{align}
and the gauge fixing parameter are governed by the differential equations:
\begin{align}
  \gamma_\xi(g_{\rm R}^2) := \mu\frac{\partial \xi_{\rm R}}{\partial\mu}  &=2\lambda_{\rm R} \xi_{\rm R} (\xi_{\rm R}-1) \left(\xi_{\rm R}-{1 \over 2}\right)
  C_2(G) \frac{g_{\rm R}^2}{(4\pi)^2}  ,
\label{RG-xi2}
  \\
  \gamma_\lambda(g_{\rm R}^2) := \mu\frac{\partial \lambda_{\rm R}}{\partial\mu}  &=2\lambda_{\rm R} \left[ \frac{13}6-\frac{\lambda_{\rm R}}{2} + \lambda_{\rm R} \xi_{\rm R} (1-\xi_{\rm R})\right]
C_2(G)  \frac{g_{\rm R}^2}{(4\pi)^2}  .
\label{RG-lambda2}
\end{align}

Eq.~(\ref{betaf2}) is a closed equation and it is solved exactly, 
\begin{equation}
  g_R^2(\mu) = {g_R^2(\mu_0) \over 1+ {22 \over 3}{C_2(G) \over (4\pi)^2}g_R^2(\mu_0) \ln {\mu \over \mu_0}} 
  = {1 \over {22 \over 3}{C_2(G) \over (4\pi)^2} \ln {\mu \over \Lambda_{\rm QCD}}}  ,
\label{betasol}
\end{equation}
where we have used the boundary condition $g_R(\mu_0)=\infty$ at $\mu_0=\Lambda_{\rm QCD}$. 
The remaining two equations (\ref{RG-xi2}) and (\ref{RG-lambda2}) are coupled equation. 
 Using the solution (\ref{betasol}) and rewriting the derivative 
${1 \over g_R^2} \mu {\partial \over \partial \mu}$
in (\ref{RG-xi2}), (\ref{RG-lambda2}) as
\begin{equation}
  {1 \over g^2} \mu {\partial \over \partial \mu}
  = {22 \over 3}{C_2(G) \over (4\pi)^2} \ln {\mu \over \Lambda_{\rm QCD}}  \mu {\partial \over \partial \mu}
  =  {22 \over 3}{C_2(G) \over (4\pi)^2}   {\partial \over \partial \ln \ln {\mu \over \Lambda_{\rm QCD}}}  ,
\label{difftransf}
\end{equation}
we can eliminate $g^2$ dependence in (\ref{RG-xi2}) and (\ref{RG-lambda2}) as 
\begin{align}
  {\partial \xi_{\rm R} \over \partial \ln \ln {\mu \over \Lambda_{\rm QCD}}}    
  &= {3 \over 11} \lambda_{\rm R} \xi_{\rm R} (\xi_{\rm R}-1) \left(\xi_{\rm R}-{1 \over 2}\right)  ,
\label{RG-xi3}
  \\
     {\partial  \lambda_{\rm R} \over \partial \ln \ln {\mu \over \Lambda_{\rm QCD}} } 
  &= {3 \over 11} \lambda_{\rm R} \left[ \frac{13}6-\frac{\lambda_{\rm R}}{2} + \lambda_{\rm R} \xi_{\rm R} (1-\xi_{\rm R})\right]    .
\label{RG-lambda3}
\end{align}
They are still coupled equation and we can not solve them  exactly. 
Therefore, we restrict our consideration to the subspace in which the flow is always confined.  It is the fixed subspace consisting of the fixed points of the RG.
The fixed point (to one-loop order) is obtained by solving the algebraic equation simultaneously: 
\begin{equation}
   \lambda_{\rm R}\xi_{\rm R}(\xi_{\rm R}-1) \left(\xi_{\rm R}-{1 \over 2}\right)  =0 ,
  \quad 
   \lambda_{\rm R} \left[ \frac{13}6-\frac{\lambda_{\rm R}}{2} + \lambda_{\rm R}\xi_{\rm R}(1-\xi_{\rm R})\right] =0 .
\end{equation}


\begin{figure}[htbp]
\begin{center}
\includegraphics[height=7cm]{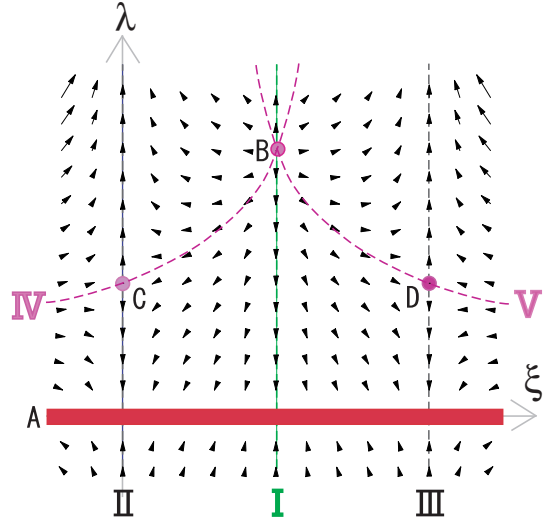}
\caption{RG flow in the $(\xi,\lambda)$ plane.}
\label{fig:flow1}
\end{center}
\end{figure}

\begin{figure}[htbp]
\begin{center}
\includegraphics[height=7cm]{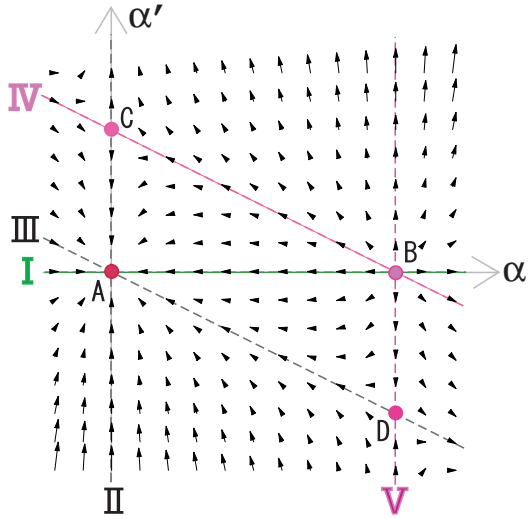}
\caption{RG flow in the $(\alpha,\alpha')$ plane.}
\label{fig:flow2}
\end{center}
\end{figure}


We find one fixed line consisting of fixed points and three isolated fixed points in the $(\xi_{\rm R},\lambda_{\rm R})$ plane,  or equivalently four isolated fixed points in the $(\alpha_{\rm R}, \alpha'_{\rm R})$ plane:
\begin{enumerate}
\item[A.] 
 The line of fixed points: $\lambda_{\rm R} =0, \xi_{\rm R} \in {\bf R}$ corresponds to an isolated fixed point $(\alpha_{\rm R}, \alpha'_{\rm R})=(0,0)$.
\item[B.]
 $(\xi_{\rm R},\lambda_{\rm R}) =({1 \over 2},{26 \over 3})$ corresponds to $(\alpha_{\rm R}, \alpha'_{\rm R})=({26 \over 3},0)$.
\item[C.]
 $(\xi,\lambda) =(0,{13 \over 3})$ corresponds to $(\alpha_{\rm R}, \alpha'_{\rm R})=(0,{13 \over 3})$.
\item[D.]
 $(\xi_{\rm R},\lambda_{\rm R}) =(1,{13 \over 3})$ corresponds to $(\alpha_{\rm R}, \alpha'_{\rm R})=({26 \over 3},-{13 \over 3})$ 
\end{enumerate}

If the two parameters $\xi_{\rm R}, \lambda_{\rm R}$ are set equal to one of the fixed points, the theory remains forever on the fixed subspace.
If the system starts from other points and the scale $\mu$ is decreased, it evolves into the infrared (IR) region according to a couple of differential equations 
(\ref{betaf2}), (\ref{RG-xi2}) and (\ref{RG-lambda2}).

Among the RG flows, the five RG flows
(I, ${\rm I\!I}$, ${\rm I\!I\!I}$, ${\rm I\!V}$, V) connecting the fixed points $A,B,C,D$ form the watershed (or backbone) in the flow diagram.
\begin{subequations}
\begin{alignat}{3}
 \text{I.} & \quad & \quad \xi_{\rm R} & ={1 \over 2} , & \quad & \alpha'_{\rm R} =0.
 \\
 \text{${\rm I\!I}$.} & & \quad \xi_{\rm R} & =0 , & & \alpha_{\rm R} =0.
 \\
 \text{${\rm I\!I\!I}$.} & & \quad \xi_{\rm R} & =1, & & \alpha'_{\rm R} =- {1 \over 2} \alpha.
 \\
 \text{${\rm I\!V}$.} & & \quad \lambda_{\rm R} & ={13 \over 3}{1 \over 1-\xi_{\rm R}}, & & \alpha'_{\rm R} =- {1 \over 2} \alpha_{\rm R} + {13 \over 3}.
 \\
 \text{V.} & & \quad \lambda_{\rm R} & ={13 \over 3}{1 \over \xi_{\rm R}}, & & \alpha_{\rm R} ={26 \over 3}.
\end{alignat}
\end{subequations}


\begin{figure}[htbp]
\begin{center}
\includegraphics[height=3cm]{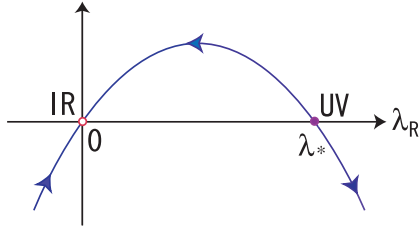}
\caption{The flow of the gauge parameter $\lambda$ in the generalized Lorentz gauge.  The arrow is directed to the IR region, i.e., decreasing the renormalization scale $\mu$.}
\label{fig:GaugeParameterFlow}
\end{center}
\end{figure}


In what follows, we use $\lambda$ to denote the initial value, $\bar\lambda:=\lambda_R$ the running parameter and $\lambda_{*}$ the fixed point of RG. 
First, we consider the fixed subspace II, i.e., $\xi_{\rm R}=0$ in which $\lambda_{*}:={13 \over 3}$ and the $\lambda_{\rm R}$ is governed by 
\begin{align}
     {\partial  \lambda_{\rm R} \over \partial \ln \ln {\mu \over \Lambda_{\rm QCD}} } 
  &= {3 \over 22} \lambda_{\rm R} \left[ \lambda_{*} - \lambda_{\rm R}    \right]  .
\label{RG-lambda4}
\end{align}
This equation is easily integrated  
\begin{align}
    \int_{\lambda}^{\bar{\lambda}}  {d  \lambda_{\rm R} \over \lambda_{\rm R} \left[ \lambda_{\rm R}  - \lambda_{*} \right]} 
  &= - {3 \over 22}  \int_{\mu_0}^{\mu} d \ln \ln {\mu \over \Lambda_{\rm QCD}}   ,
\label{RG-lambda5}
\end{align}
or
\begin{align}
    \int_{\lambda}^{\bar{\lambda}}  d  \lambda_{\rm R} \left[ {1 \over  \lambda_{\rm R} } - {1 \over \lambda_{\rm R}  - \lambda_{*}  } \right]
  &=   {13 \over 22}  \int_{\mu_0}^{\mu} d \ln \ln {\mu \over \Lambda_{\rm QCD}}    ,
\label{RG-lambda55}
\end{align}
to yield  
\begin{align}
   \bar{\lambda} = \lambda_{*} \left\{ 
 1 - {\lambda-\lambda_{*} \over \lambda} 
 \left[  {\ln (\mu/\Lambda_{\rm QCD})  \over \ln (\mu_0/\Lambda_{\rm QCD}) } \right]^{-{13 \over 22} }
 \right\}^{-1}  
= \lambda_{*} \left\{ 
 1 - {\lambda-\lambda_{*} \over \lambda} 
 \left(  {\bar{g}^2 \over g^2} \right)^{+{13 \over 22} }
 \right\}^{-1}   .
\label{RG-lambda6}
\end{align}
Note that $\bar{\lambda}=0$ for $\lambda=0$ and hence $\bar{\lambda}=0$ is an IR fixed point, while the point $\bar{\lambda}=\lambda_{*}$ is an UV fixed point.
$\bar{\lambda}=\lambda_{*}$ for $\lambda=\lambda_{*}$. For other initial values of $\lambda$, we find that $\bar{\lambda} \downarrow \lambda_{*}$ for $\lambda>\lambda_{*}$ and $\bar{\lambda} \uparrow \lambda_{*}$ for $\lambda_{*}>\lambda>0$,  
while 
$\bar{\lambda} \rightarrow -\infty$ for $\lambda<0$.

The anomalous dimension of the gluon field is $\xi$-independent and is given by 
\begin{equation}
 \gamma_A(g_{\rm R}^2)  
 =- \frac{1}{2} \left( \lambda_{*}- \lambda_{\rm R} \right) C_2(G)
  \frac{g_{\rm R}^2}{(4\pi)^2}  ,
\label{adimgluon}
\end{equation}
and the ghost field has the (also $\xi$-independent) anomalous dimension
\begin{equation}
 \gamma_C(g_{\rm R}^2) 
 =- \frac{3-\lambda_{\rm R}}4 C_2(G)
  \frac{g_{\rm R}^2}{(4\pi)^2} .
\label{adimghost}  
\end{equation}
Substituting (\ref{RG-lambda6}) for $\bar{\lambda} \equiv \lambda_{\rm R}$ into (\ref{adimgluon}) and (\ref{adimghost}), 
we can regard the anomalous dimensions as the functions of the gauge coupling $g^2$ alone.

For the $\lambda$ in the neighborhood of $\lambda_{*}$, (\ref{RG-lambda6}) reads 
\begin{align}
   \bar{\lambda} = \lambda_{*} + \delta \lambda  
 \left(  {\bar{g}^2 \over g^2} \right)^{{13 \over 22} }
+ O((\delta \lambda)^2) , 
\quad \delta \lambda := \lambda-\lambda_{*} .
\label{RG-lambda7}
\end{align}
Therefore, the exponential factor for the gluon reads
\begin{align}
 & \exp \left\{ 2 \int_{g^2}^{\bar{g}^2}dg_R^2 {\gamma_A(g_R^2) \over \beta(g_R^2)} \right\} 
 =    \exp \left\{ \int_{g^2}^{\bar{g}^2}dg_R^2 {1 \over {11 \over 3}g_R^2 }  \frac{1}{2} \left[ - \delta\lambda 
\left(  {g_R^2 \over g^2} \right)^{{13 \over 22} }
 \right]  
  + O(\delta\lambda^2) \right\} 
\nonumber\\
=&   \exp \left\{
  -    \delta\lambda {3 \over 22}{22 \over 13}
\left[ \left( {\bar{g}^2 \over g^2} \right)^{{13 \over 22}} -1 \right]   + O(\delta\lambda^2) \right\} 
\nonumber\\
=&  \left\{
 1 -  {\delta\lambda \over \lambda_{*}}
\left[ \left( {\bar{g}^2 \over g^2} \right)^{{13 \over 22}} -1 \right]   + O(\delta\lambda^2) \right\} 
\nonumber\\
=& {\lambda \over \lambda_{*}} 
-   {\delta\lambda \over \lambda_{*}} \left( {\bar{g}^2 \over g^2} \right)^{{13 \over 22} }  
+ O(\delta\lambda^2)  
\rightarrow   {\lambda \over \lambda_{*}} 
 \quad (\bar{g} \rightarrow 0) ,
\label{gluonf1}
\end{align}
where we have used
$
1 + {\delta\lambda \over \lambda_{*}}
= 1+ {\lambda- \lambda_{*} \over \lambda_{*}} 
= {\lambda \over \lambda_{*}} 
$.
For the ghost, the factor reads
\begin{align}
 & \exp \left\{ 2 \int_{g^2}^{\bar{g}^2}dg_R^2 {\gamma_C(g_R^2) \over \beta(g_R^2)} \right\} 
 =    \exp \left\{ \int_{g^2}^{\bar{g}^2}dg_R^2 {1 \over {11 \over 3}g_R^2 }  \frac{1}{4} \left[ 3 -  \lambda_{*} - \delta\lambda 
\left(  {g_R^2 \over g^2} \right)^{{13 \over 22} }
 \right]  
  + O(\delta\lambda^2) \right\} 
\nonumber\\
=&  \left( {\bar{g}^2 \over g^2} \right)^{{3(3-\lambda_{*}) \over 44} } \exp \left\{
  -    \delta\lambda {3 \over 44}{22 \over 13}
\left[ \left( {\bar{g}^2 \over g^2} \right)^{{13 \over 22}} -1 \right]   + O(\delta\lambda^2) \right\} 
\nonumber\\
=&  \left( {\bar{g}^2 \over g^2} \right)^{-{1 \over 11} } \left\{
 1 -   {\delta\lambda \over 2\lambda_{*}}
\left[ \left( {\bar{g}^2 \over g^2} \right)^{{13 \over 22}} -1 \right]    \right\} + O(\delta\lambda^2)
\nonumber\\
=&  
 \left( 1 + {\delta\lambda \over 2\lambda_{*}} \right) \left( {\bar{g}^2 \over g^2} \right)^{-{1 \over 11}} 
-   {\delta\lambda \over 2\lambda_{*}} \left( {\bar{g}^2 \over g^2} \right)^{{11 \over 22} }  + O(\delta\lambda^2) 
 \rightarrow \infty .
\label{ghostf1}
\end{align}
The first term diverges as $\bar{g} \rightarrow 0$. 
In this case the superconvergence relation does not exist. 
In this case, the asymptotic form of the ghost form factor is given by
\begin{align}
  G(k^2) \cong  \left( 1 + {\delta\lambda \over 2\lambda_{*}} \right) \left( \ln {k^2 \over \mu^2} \right)^{{1 \over 11}}
\rightarrow \infty \quad \text{as} \quad k^2 \rightarrow \infty .
\end{align}
The ghost spectral function has the asymptotic form
\begin{align}
  \rho_{FP}(k^2) \cong -(k^2)^{-1} \left( 1 + {\delta\lambda \over 2\lambda_{*}} \right) \left( \ln {k^2 \over \mu^2} \right)^{{1 \over 11}-1} .
\end{align}
These results agree with those of \cite{Xu96}.

For sufficiently small $\lambda$ in the neighborhood of $\lambda=0$, (\ref{RG-lambda6}) reads  
\begin{align}
   \bar{\lambda} \cong \lambda  
 \left(  {\bar{g}^2 \over g^2} \right)^{-{13 \over 22} } .
\label{RG-lambda8}
\end{align}
Therefore, the exponential factor for the gluon reads
\begin{align}
 & \exp \left\{ 2 \int_{g^2}^{\bar{g}^2}dg_R^2 {\gamma_A(g_R^2) \over \beta(g_R^2)} \right\} 
 =    \exp \left\{ \int_{g^2}^{\bar{g}^2}dg_R^2 {1 \over {11 \over 3}g_R^2 }  \frac{1}{2} \left[ \lambda_{*} -  \lambda 
\left(  {g_R^2 \over g^2} \right)^{-{13 \over 22} }
 \right]  
  \right\} 
\nonumber\\
=&  \left( {\bar{g}^2 \over g^2} \right)^{{13 \over 22} } \exp \left\{
  {\lambda \over \lambda_{*}}
\left[ \left( {\bar{g}^2 \over g^2} \right)^{-{13 \over 22}} -1 \right]   \right\} 
\nonumber\\
=&  \left( {\bar{g}^2 \over g^2} \right)^{{13 \over 22} } \left\{
 1+ {\lambda \over \lambda_{*}}
\left[ \left( {\bar{g}^2 \over g^2} \right)^{-{13 \over 22}} -1 \right]   + O(\lambda^2) \right\} 
\nonumber\\
=& {\lambda \over \lambda_{*}} 
+ \left( 1- {\lambda \over \lambda_{*}}  \right)   \left( {\bar{g}^2 \over g^2} \right)^{{13 \over 22} }  
\rightarrow {\lambda \over \lambda_{*}} 
 \quad (\bar{g} \rightarrow 0) .
\label{gluonf2}
\end{align}
This result is the same as the above result (\ref{gluonf1}). 
For the ghost, however, the factor reads
\begin{align}
 & \exp \left\{ 2 \int_{g^2}^{\bar{g}^2}dg_R^2 {\gamma_C(g_R^2) \over \beta(g_R^2)} \right\} 
 =    \exp \left\{ \int_{g^2}^{\bar{g}^2}dg_R^2 {1 \over {11 \over 3}g_R^2 }  \frac{1}{4} \left[ 3 -  \lambda 
\left(  {g_R^2 \over g^2} \right)^{-{13 \over 22} }
 \right]  
  \right\} 
\nonumber\\
=&  \left( {\bar{g}^2 \over g^2} \right)^{{9 \over 44} } \exp \left\{
  {\lambda \over 2\lambda_{*}}
\left[ \left( {\bar{g}^2 \over g^2} \right)^{-{13 \over 22}} -1 \right]   \right\} 
\nonumber\\
=&  \left( {\bar{g}^2 \over g^2} \right)^{{9 \over 44} } \left\{
 1+ {\lambda \over 2\lambda_{*}}
\left[ \left( {\bar{g}^2 \over g^2} \right)^{-{13 \over 22}} -1 \right]   + O(\lambda^2) \right\} 
\nonumber\\
=&  {\lambda \over 2\lambda_{*}}
  \left( {\bar{g}^2 \over g^2} \right)^{-{17 \over 44} } 
+ \left( 1 -  {\lambda \over 2\lambda_{*}} \right) \left( {\bar{g}^2 \over g^2} \right)^{{9 \over 44}} 
 \rightarrow \infty .
\label{ghostf2}
\end{align}
This factor diverges for non-zero $\lambda$ as $\bar{g} \rightarrow 0$ no matter how small $\lambda$ is chosen. 
In this case the superconvergence relation does not exist. 
Only for an initial value $\lambda=0$, this factor goes to zero as $\bar{g} \rightarrow 0$ and the superconvergence relation holds. 
Therefore, there is a discontinuity at $\lambda=0$ for the ghost factor.  For the superconvergence to be realized for the ghost propagator, we need a fine tuning of $\lambda$.
In this case, the asymptotic form of the ghost form factor is given by
\begin{align}
  G(k^2) \cong  {\lambda \over 2\lambda_{*}} \left( \ln {k^2 \over \mu^2} \right)^{{17 \over 44}} 
\rightarrow \infty \quad \text{as} \quad k^2 \rightarrow \infty .
\end{align}
The ghost spectral function has the asymptotic form
\begin{align}
  \rho_{FP}(k^2) \cong -(k^2)^{-1}  {\lambda \over 2\lambda_{*}}
 \left( \ln {k^2 \over \mu^2} \right)^{{17 \over 44}-1} .
\end{align}
This case has not been worked out in \cite{Xu96}.

Next, in the fixed subspace I, i.e., $\xi=1/2$, the final result is obtained by replacing ${13 \over 22}$ by ${13 \over 44}$ with the identification $\lambda_{*}={26 \over 3}$. 
\begin{align}
   \bar{\lambda} = \lambda_{*} \left\{ 
 1 - {\lambda-\lambda_{*} \over \lambda} 
 \left[  {\ln (\mu/\Lambda_{\rm QCD})  \over \ln (\mu_0/\Lambda_{\rm QCD}) } \right]^{-{13 \over 44} }
 \right\}^{-1}  
= \lambda_{*} \left\{ 
 1 - {\lambda-\lambda_{*} \over \lambda} 
 \left(  {\bar{g}^2 \over g^2} \right)^{+{13 \over 44} }
 \right\}^{-1}   .
\label{RG-lambda61}
\end{align}
The results for the $\lambda$ in the neighborhood of $\lambda_{*}$ are obtained by replacing  the exponent 
${1 \over 11}$ by ${17 \over 44}$.
The results for sufficiently small $\lambda$ in the neighborhood of $\lambda=0$ are obtained by replacing the exponent ${17 \over 44}$   by ${1 \over 11}$.
Finally, in the fixed subspace III, i.e., $\xi=1$, the same result as I is obtained, since the theory has the reflection symmetry $\xi \leftrightarrow 1-\xi$. 
They are new results for the generalized Landau gauge. 

Thus the ghost can escape from the superconvergence except for the Landau gauge $\lambda=0$.  
This point is very important for our result due to general principle to match the results of approximate solutions of the Schwinger--Dyson equation for gluodynamics.  
The Landau gauge limit $\lambda \downarrow 0$ may be rather subtle.

\subsection{Modified MA gauge}

We discuss only the case $G=SU(2)$.
It was shown \cite{Schaden99,SIK01,KS01,EW02} that the RG equation for the gauge fixing parameter $\beta$ of the diagonal part is given by \cite{KS01,SIK01}
\begin{equation}
\mu\frac{\partial}{\partial\mu} \beta_{\rm R}
 =\frac{44}3 \beta_{\rm R}\frac{g_{\rm R}^2}{(4\pi)^2} ,
\label{MAbeta}
\end{equation}
and the gauge fixing parameter $\alpha$ of the off-diagonal part obeys the RG equation \cite{Schaden99,KS01,SIK01}
\begin{equation}
\mu\frac{\partial}{\partial\mu} \alpha_{\rm R}
 =
  \left[-2\alpha_{\rm R}^2+\frac83\alpha_{\rm R}-6
  \right] \frac{g_{\rm R}^2}{(4\pi)^2} .
\label{MAalpha}
\end{equation}

We apply  (\ref{difftransf}) to rewrite the differential equation (\ref{MAbeta}) into the form which does not explicitly depend on $g^2$: \begin{align}
{22 \over 3} 2   {\partial \over \partial \ln \ln {\mu \over \Lambda_{\rm QCD}}}  \beta_{\rm R}
 =\frac{44}3  \beta_{\rm R} ,
\end{align}
which is easily solved:

\begin{align}
 \int_{\beta}^{\bar{\beta}} {d \beta \over \beta} = \int_{\mu_0}^{\mu} d \ln \ln {\mu \over \Lambda_{\rm QCD}} ,
\end{align}
and
\begin{align}
  \bar{\beta} =  \beta {\ln (\mu/\Lambda_{\rm QCD})  \over \ln (\mu_0/\Lambda_{\rm QCD}) } 
=  \beta  \left(  {\bar{g}^2 \over g^2} \right)^{-1}
=  \beta    {g^2 \over \bar{g}^2} .
\end{align}
As $\mu \rightarrow \infty$ or $\bar{g} \rightarrow 0$, $\bar{\beta} \rightarrow + \infty$ for $\beta>0$ and 
$\bar{\beta} \rightarrow -\infty$ for $\beta<0$, while $\bar{\beta}\equiv 0$ for $\beta=0$.


\begin{figure}[htbp]
\begin{center}
\includegraphics[height=4cm]{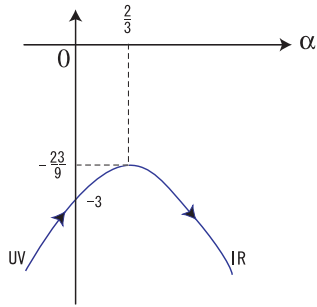}
\caption{The flow of the gauge parameter $\alpha$ in the modified MA gauge.  The arrow is directed to the IR region, i.e., decreasing the renormalization scale $\mu$.}
\label{fig:GaugeParameterFlowMA}
\end{center}
\end{figure}


In the similar way, (\ref{MAalpha}) is cast into 
\begin{align}
 {22 \over 3} 2   {\partial \over \partial \ln \ln {\mu \over \Lambda_{\rm QCD}}} \alpha_{\rm R}
 =
  \left[ -2\alpha_{\rm R}^2+\frac83\alpha_{\rm R}-6
  \right]   ,
\label{MAalpha2}
\end{align}
or
\begin{align}
   \int_{\alpha}^{\bar{\alpha}} {d\alpha_{\rm R} \over -2\alpha_{\rm R}^2+\frac83\alpha_{\rm R}-6}
 = {3 \over 44} \int_{\mu_0}^{\mu} d \ln \ln {\mu \over \Lambda_{\rm QCD}}  .
\label{MAalpha3}
\end{align}
First, we consider sufficiently small $\alpha$, neglecting the order $\alpha^2$ term, 
\begin{align}
   \int_{\alpha}^{\bar{\alpha}} {d\alpha_{\rm R} \over \frac83\alpha_{\rm R}-6}
 = {3 \over 44} \int_{\mu_0}^{\mu} d \ln \ln {\mu \over \Lambda_{\rm QCD}}  ,
\end{align}
which yields
\begin{align}
   \bar{\alpha}(\mu) = {9 \over 4} + \left( \alpha - {9 \over 4} \right) 
\left[ {\ln (\mu/\Lambda_{\rm QCD}) \over \ln (\mu_0/\Lambda_{\rm QCD}) } \right]^{{2 \over 11}} 
= {9 \over 4} + \left( \alpha - {9 \over 4} \right)  \left( {\bar{g}^2 \over g^2} \right)^{-{2 \over 11} } 
\rightarrow -\infty \quad (\mu \rightarrow \infty) .
\end{align}
Next, we take into account the $O(\alpha^2)$ term too. 
Applying the formula
\begin{align}
  \int {dx \over ax^2+bx+c} = {2 \over \sqrt{4ac-b^2}} \arctan {2ax+b \over \sqrt{4ac-b^2}} \quad (b^2<4ac) ,
\end{align}
to (\ref{MAalpha3}), 
we obtain
\begin{align}
   \arctan {-3\bar{\alpha}+2 \over \sqrt{23}} 
= \arctan {-3\alpha+2 \over \sqrt{23}} 
+ {\sqrt{23} \over 22} \ln \left[ {\ln (\mu/\Lambda_{\rm QCD})  \over \ln (\mu_0/\Lambda_{\rm QCD}) } \right] .
\end{align}
Thus the running gauge parameter obeys 
\begin{align}
 \bar{\alpha}(\mu) =& {2 \over 3} - {\sqrt{23} \over 3} 
 \tan \left\{ \arctan {-3\alpha+2 \over \sqrt{23}} 
+ {\sqrt{23} \over 22} \ln \left[ {\ln (\mu/\Lambda_{\rm QCD})  \over \ln (\mu_0/\Lambda_{\rm QCD}) } \right] \right\} 
\\
=& {2 \over 3} - {\sqrt{23} \over 3} 
 \tan \left\{ \arctan {-3\alpha+2 \over \sqrt{23}} 
+ {\sqrt{23} \over 22} \ln \left[ {g^2  \over  \bar{g}^2} \right] \right\} .
\end{align}
This shows that $\bar{\alpha} \rightarrow -\infty$ as $\mu \rightarrow \infty$ irrespective of the value of $\alpha$. 
Note that $\arctan x$ has multivalued, unless $-\pi/2<\arctan x<\pi/2$. 

The anomalous dimension of the diagonal gluon is given by \cite{SIK01}
\begin{equation}
\gamma_a(g_{\rm R}^2)
 =-\frac{22}3\frac{g_{\rm R}^2}{(4\pi)^2 },
\end{equation}
while the off-diagonal gluons have the anomalous dimension \cite{Schaden99,SIK01}
\begin{equation}
\gamma_A(g_{\rm R}^2)
 =-\frac{g_{\rm R}^2}{(4\pi)^2} 
   \left[\frac{17}6-\frac{\alpha_{\rm R}}2-\beta_{\rm R}
   \right] .
\end{equation}
The anomalous dimensions of the diagonal ghost and diagonal antighost is calculated as \cite{SIK01}
\begin{equation}
\gamma_c(g_{\rm R}^2) = - \gamma_{\bar{c}}(g_{\rm R}^2)
 =-(3+\alpha_{\rm R})
   \frac{g_{\rm R}^2}{(4\pi)^2},
\end{equation}
while the off-diagonal ghosts read \cite{SIK01}
\begin{equation}
\gamma_C(g_{\rm R}^2)
 =-\frac{g_{\rm R}^2}{(4\pi)^2}
  (3-\beta_{\rm R}) .
\end{equation}
For the derivation of anomalous dimensions, see  \cite{SIK01,KS01}.

For the diagonal gluon, the correction factor reads
\begin{align}
 & \exp \left\{ 2 \int_{g^2}^{\bar{g}^2}dg_R^2 {\gamma_a(g_R^2) \over \beta(g_R^2)} \right\} 
 =    \exp \left\{ \int_{g^2}^{\bar{g}^2}dg_R^2 {1 \over g_R^2}  
  \right\} 
=    {\bar{g}^2 \over g^2}  
= \left[ {\ln (p/\Lambda_{\rm QCD})  \over \ln (\mu_0/\Lambda_{\rm QCD}) } \right]^{-1}
 \rightarrow 0 .
\end{align}
The superconvergence relation holds for the diagonal gluon.

For the off-diagonal gluon,
\begin{align}
 & \exp \left\{ 2 \int_{g^2}^{\bar{g}^2}dg_R^2 {\gamma_A(g_R^2) \over \beta(g_R^2)} \right\} 
 =    \exp \left\{ \int_{g^2}^{\bar{g}^2}dg_R^2 {1 \over {11 \over 3}g_R^2 } 
\left[\frac{17}6-\frac{\alpha_{\rm R}}2-\beta_{\rm R}
   \right]  
  \right\} 
\nonumber\\
 =&    \exp \left\{ {3 \over 11} \left( {17 \over 6}-{1 \over 3} \right) \int_{g^2}^{\bar{g}^2}{dg_R^2 \over g_R^2} 
 - {3 \over 11}  {\sqrt{23} \over 6} \int_{g^2}^{\bar{g}^2}{dg_R^2 \over g_R^2}
\left[ \tan \left( 
{\sqrt{23} \over 22} \ln  {g^2  \over  g_R^2} 
+ \arctan {-3\alpha+2 \over \sqrt{23}} 
 \right)
   \right]    \right\} 
\nonumber\\ & \times 
\exp \left\{ -  {3 \over 11} \beta g^2 \int_{g^2}^{\bar{g}^2}{dg_R^2 \over (g_R^2)^2}  
  \right\} 
\nonumber\\
 =&  \exp \left\{ {15 \over 22} \int_{g^2}^{\bar{g}^2}{dg_R^2 \over g_R^2} 
 -  \int_{g_R^2=g^2}^{g_R^2=\bar{g}^2}d x
\left[ \tan \left( x + \arctan {-3\alpha+2 \over \sqrt{23}} 
   \right)
   \right]    \right\} 
\nonumber\\ & \times 
\exp \left\{ -  {3 \over 11} \beta g^2 \int_{g^2}^{\bar{g}^2}{dg_R^2 \over (g_R^2)^2}  
  \right\} 
\nonumber\\
=&   \left( {\bar{g}^2 \over g^2} \right)^{{15 \over 22} } 
 \left|\cos \left[ {\sqrt{23} \over 22} \ln  {g^2  \over  \bar{g}^2}
+ \arctan {-3\alpha+2 \over \sqrt{23}}  \right] \right|  \left|\cos \left[ \arctan {-3\alpha+2 \over \sqrt{23}} \right] \right|^{-1} 
\nonumber\\ & \times 
   \exp \left\{  {3 \over 11} \beta 
\left[ {g^2 \over \bar{g}^2} - 1 \right]   \right\} 
\rightarrow 0 \quad \text{for} \quad \beta \le 0
 \quad (\text{as} \quad \bar{g} \rightarrow 0) ,
\end{align}
where we have used
\begin{align}
  \int dx \tan (cx) = -{1 \over c} \ln |\cos (cx)| .
\end{align}
The RHS oscillates, but goes to zero for arbitrary $\alpha$ for $\beta \le 0$, while it goes to infinity for $\beta>0$.
The superconvergence relation holds for $\beta \le 0$.

For the off-diagonal ghost, 
\begin{align}
 & \exp \left\{ 2 \int_{g^2}^{\bar{g}^2}dg_R^2 {\gamma_c(g_R^2) \over \beta(g_R^2)} \right\} 
 =    \exp \left\{ \int_{g^2}^{\bar{g}^2}dg_R^2 {1 \over {11 \over 3}g_R^2 } 
(3-\beta_{\rm R}),
  \right\} 
\nonumber\\
 =&    \exp \left\{ {9 \over 11} \int_{g^2}^{\bar{g}^2}{dg_R^2 \over g_R^2} -  {3 \over 11} \beta g^2 \int_{g^2}^{\bar{g}^2}{dg_R^2 \over (g_R^2)^2}  
  \right\} 
\nonumber\\
=&  \left( {\bar{g}^2 \over g^2} \right)^{{9 \over 11} } 
 \exp \left\{   {3 \over 11} \beta 
\left[ {g^2 \over \bar{g}^2} - 1 \right]   \right\} 
\rightarrow 0 \quad \text{for} \quad \beta \le 0
 \quad (\text{as} \quad \bar{g} \rightarrow 0) .
\end{align}
As $\mu \rightarrow \infty$ $(\bar{g} \rightarrow 0)$, the RHS goes to zero for arbitrary $\alpha$ for $\beta \le 0$, while it goes to infinity for $\beta>0$.
The superconvergence relation holds for $\beta \le 0$.

For the diagonal ghost, the anomalous dimension of the diagonal ghost has the same magnitude as that of the diagonal antighost with opposite sign, i.e., 
$\gamma_c(g_R^2)=-\gamma_{\bar{c}}(g_R^2)$ and hence the corrections cancel, 
\begin{align}
 & \exp \left\{  \int_{g^2}^{\bar{g}^2}dg_R^2 {\gamma_c(g_R^2)+\gamma_{\bar{c}}(g_R^2) \over \beta(g_R^2)} \right\} 
 =   1 .
\end{align}
There is no correction to the tree form for the diagonal ghost--antighost propagator. 
Therefore, the superconvergence does not occur for diagonal ghost and antighost.


\end{document}